\documentclass{aa}
\usepackage{lscape}
\usepackage{amsmath,amsfonts,amssymb}
\usepackage{graphicx}
\usepackage{epstopdf}
\usepackage[]{natbib}
\bibpunct{(}{)}{;}{a}{}{,} 
\usepackage{txfonts}
\usepackage{hyperref}
\newcommand{\kms}{km\,s$^{-1}$}

\def\Mo{M_{\odot}}

\def\Ro{R_{\odot}}

\begin{document}

   \title{Improving the surface brightness-color relation for early-type stars using optical interferometry \thanks{Partly based on VEGA/CHARA observations}}

   \subtitle{}
   \titlerunning{The Araucaria Project: Improving the surface brightness-color relation for early-type stars}

   \authorrunning{M. Challouf et al.}

 \author{M. Challouf \inst{1,2}, N. Nardetto \inst{1}, D. Mourard \inst{1}, D. Graczyk \inst{3}, H. Aroui \inst{2}, O. Chesneau\inst{1}, O. Delaa \inst{1}, G. Pietrzy\'nski \inst{3,4}, W. Gieren \inst{3}, R. Ligi \inst{1}, A. Meilland \inst{1}, K. Perraut \inst{5}, I. Tallon-Bosc\inst{6}, H. McAlister\inst{7,8}, T. ten~Brummelaar\inst{8}, J. Sturmann\inst{8}, L. Sturmann\inst{8}, N. Turner\inst{8}, C. Farrington\inst{8}, N. Vargas \inst{8}, and N. Scott \inst{8}}


 \institute{Laboratoire Lagrange, UMR7293, UNS/CNRS/OCA, 06300 Nice, France \and Laboratoire Dynamique Mol\'eculaire et Mat\'eriaux Photoniques, UR11ES03, Universit\'e de Tunis/ESSTT, Tunisie \and Departamento de Astronom\'ia, Universidad de Concepci\'on, Casilla 160-C, Concepci\'on \and Warsaw University Observatory, AL. Ujazdowskie 4, 00-478, Warsaw, Poland \and Institut d$^{\prime}$Astrophysique et de Plan\'etologie de Grenoble, CNRS-UJF UMR 5571, 414 rue de la Piscine, 38400 St Martin d$^{\prime}$H\`eres, France  \and Universit\'e de Lyon, 69003 Lyon, France; Universit\'e Lyon 1, Observatoire de Lyon, 9 avenue Charles Andr\'e, 69230 Saint Genis Laval, France; CNRS/UMR 5574, Centre de Recherche Astroph. de Lyon; Ecole Normale Sup\'erieure, 69007 Lyon, France \and Georgia State University, P.O. Box 3969, Atlanta GA 30302-3969, USA  \and CHARA Array, Mount Wilson Observatory, 91023 Mount Wilson CA, USA}

   \date{Received;accepted}


  \abstract
   {The method of distance determination of eclipsing binaries consists of combining  the radii of both components determined from spectro-photometric observations with their respective angular diameters derived from the surface brightness-color relation (SBC). However, the largest limitation of the method comes from the uncertainty on the SBC relation: about 2\% for late-type stars (or 0.04 magnitude) and more than 10\% for early-type stars (or 0.2 magnitude).}
   { The aim of this work is to improve the SBC relation for early-type stars in the  $-1 \leq V-K \leq 0$ color domain, using optical interferometry.}
   {Observations of eight B- and A-type stars were secured with the VEGA/CHARA instrument in the visible. The derived uniform disk angular diameters were converted into limb darkened angular diameters and included in a larger sample of 24 stars, already observed by interferometry, in order to derive a revised empirical relation for O, B, A spectral type stars with a V-K color index ranging from -1 to 0. We also took the opportunity to check the consistency of the SBC relation up to $V-K \simeq 4$ using 100 additional measurements. }
   {We determined the uniform disk angular diameter for the eight following stars: $\gamma$~Ori, $\zeta$~Per, $8$~Cyg, $\iota$~Her, $\lambda$~Aql, $\zeta$~Peg, $\gamma$~Lyr, and $\delta$~Cyg with V-K color ranging from -0.70 to 0.02 and typical precision of about $1.5\%$. Using our total sample of 132 stars with $V-K$ colors index ranging from about $-1$ to $4$, we provide a revised SBC relation. For late-type stars ($0 \leq V-K \leq 4$), the results are consistent with previous studies. For early-type stars ($-1 \leq V-K \leq 0$), our new VEGA/CHARA measurements combined with a careful selection of the stars (rejecting stars with environment or stars with a strong variability), allows us to reach an unprecedented precision of about 0.16 magnitude or $\simeq 7\%$ in terms  of angular diameter.}
 {We derive for the first time a SBC relation for stars between O9 and A3, which provides a new and reliable tool for the distance scale calibration.}


   \maketitle
%

\section{Introduction}

The distance measurements to extragalactic targets in the last century revolutionized our understanding of the distance scale of the universe. The distance to the LMC is a critical rung on the cosmic distance ladder, and numerous independent methods involving, for instance, RR Lyrae stars \citep{1997MNRAS.284..761F,2008AJ....136..272S,2008AJ....135.1993P}, Cepheids \citep{1985ApJ...296..169B,1991ApJ...372..597E,1992ApJ...389..657E,1996ASPC...88....9F,2008ApJ...679...71F}, or red clump stars \citep{1998AcA....48....1U,1998ApJ...509L..25U,2002AJ....124.2633P,2012MNRAS.419.1637L} have been used to derive its distance.

The main goal of the long term program called the Araucaria project is to significantly improve the calibration of the cosmic distance scale based on observations of several distance indicators in nearby galaxies \citep{2005ApJ...628..695G}. Eclipsing binary systems are particularly important to provide the zero point of the extragalactic distances and study in detail populational dependence on other distance indicators like RR Lyrae stars, Cepheids, red clump stars, etc. Thirteen long period systems composed of late-type giants were analyzed in the LMC and SMC so far: eight in the  LMC \citep{2009ApJ...697..862P, 2013Natur.495...76P}, and five in the SMC \citep{2012ApJ...750..144G, 2014ApJ...780...59G}. For such systems, the linear dimension of both components can be measured with a precision  up to of 1\% from the analysis of high-quality spectroscopic and photometric data (e.g., \citet{2010A&ARv..18...67T}). The distance to an eclipsing binary follows from the dimensions determined in this way, plus the angular diameters derived from the absolute surface brightness, which is very well calibrated for late-type stars  \citep{2005MNRAS.357..174D}. This conceptually  very simple technique very weakly depends on reddening and metallicity, and provides the most accurate tool for measuring distances to nearby galaxies \citep{2013Natur.495...76P,2014ApJ...780...59G}.

However the heart of this method, the surface brightness-color relation, is very well calibrated only for late-type stars which significantly limits its usage. The late-type systems composed of main-sequence stars are usually faint,  while those composed of giants have very long periods (several hundred days) that makes them very difficult to find. As a result, only about 45 late-type systems, well suited to precise distance determination have been discovered so far in the Magellanic Clouds by the Optical Gravitational Microlensing Experiment (OGLE) \citep{2013AcA....63..323P, 2011AcA....61..103G}.
On the other hand, many more relatively bright systems are known in nearby galaxies \citep{2013ApJ...768....6M, 2011AcA....61..103G,2003AcA....53....1W,2004AcA....54....1W,2006ApJ...652..313B,2001AJ....121..870M,2001AJ....122.1383M, 2006A&A...459..321V,2013AcA....63..323P}. Therefore, in order to derive the distance to nearby galaxies and to study the geometry of the Magellanic Clouds, it is imperative to calibrate surface brightness-color relation for early-type stars.

The purpose of this paper is to improve the SBC relation for early-type stars by using the resolving power of the Visible spEctroGraph and polArimeter (VEGA) beam combiner \citep{2009A&A...508.1073M} operating at the focus of the CHARA (The Center for High Angular Resolution Astronomy) Array \citep{2005ApJ...628..453T} located at Mount Wilson Observatory (California, USA). The CHARA array consists of six telescopes of 1 meter in diameter, configured in a Y shape, which offers 15 different baselines from 34 meters to 331 meters. These baselines can achieve a spatial resolution up to 0.3~mas in the visible which is necessary in order to resolve early-type stars. Early-type stars are very small in terms of angular diameter and can be affected by several physical phenomena, like fast rotation, winds, and environment, which can potentially bias the interferometric measurements.

\begin{sidewaystable*}
\begin{center}
\caption[]{Physical parameters of the stars in our sample}
\label{Tab1}
\setlength{\doublerulesep}{\arrayrulewidth}
\begin{tabular}{p{2cm}|cccccccc}
\hline \noalign{\smallskip} \hline\hline\hline\hline

               & $\lambda$ Aql        & $\gamma$ Lyr        & $\gamma$ Ori             & $8$ Cyg          & $\iota$ Her         & $\zeta$ Per        & $\zeta$ Peg     & $\delta$ Cyg \\
               & HD 177756            & HD 176437           & HD 35468                 & HD 184171        & HD 160762           & HD 24398           & HD 214923       & HD 186882    \\
\hline\hline \noalign{\smallskip}
RA             & 19:06:14             &  18:58:56           &  05:25:07                &  19:31:46        & 17:39:27             & 03:54:07          & 22:41:27        &  19:44:58     \\
Dec      & $+04^{\circ}52'57''$ &$+32^{\circ}41'22''$ &$+06^{\circ}20'58''$&$+34^{\circ}27'10''$&$+46^{\circ}00'22''$&$+31^{\circ}53'01''$ &$+10^{\circ}49'52''$&$+45^{\circ}07'50''$  \\
S. Type $^{a}$& B9V                  & A1III               & B2III                    &   B3IV           &    B3IV              & B1Ib              & B9IV            & A0IV           \\
m$_\mathrm{V}[\mathrm{mag}]$ $^{b}$  & $3.430$              & $3.248$             & $1.637$                  &   $4.740$        &   $3.794$           & $2.850$      & $3.406$             & $2.868$ \\
m$_\mathrm{K}[\mathrm{mag}]$ $^{b}$  & $3.670$              & $3.240$             & $2.340$                  &   $5.114$        &   $4.228$           & $2.670$         & $3.565$             & $2.810$ \\
$\pi[\mathrm{mas}]$ $^{c}$& $26.37\pm0.64$   & $5.26\pm0.27$ & $12.92\pm0.52$         &   $3.79\pm0.16$  &   $7.17\pm0.13$      &$4.34\pm0.19$      & $15.96\pm0.19$  & $19.77\pm0.48$  \\
$A_{V}$ $^{d}$      & $0.03\pm0.01$        & $0.12\pm0.03$   & $0.00\pm0.06$          &   $0.15\pm0.11$&   $0.04\pm0.03$    & $0.87\pm0.05$   & $0.05\pm0.01$ & $0.04\pm0.01$  \\
$E(V-K)$ $^{e}$     & $0.03\pm0.01$        &$0.11\pm0.03$    & $0.00\pm0.05$          &  $0.12\pm0.10$ &   $0.03\pm0.03$    &$0.77\pm0.05$   & $0.04\pm0.01$ & $0.04\pm0.01$  \\
\hline \noalign{\smallskip}
$T_{\mathrm{eff}}[K]$& $11780$ $^{a}$ & $10000$ $^{a}$  & $21840$ $^{a}$   & $16100$ $^{m}$& $17000$ $^{m}$  & $22580$ $^{a}$& $11430$ $^{a}$& $10150$ $^{a}$\\
$M[M_{\odot}]$ & $2.99$ $^{j}$ & $5.76$ $^{k}$  & $7.77$ $^{j}$   & $6.40$ $^{m}$ & $6.30$ $^{m}$ & $15.50$ $^{l}$& $3.22$ $^{j}$ & $2.93$ $^{j}$      \\
$R[R_{\odot}]$ & $2.24$ $^{j}$ & $15.40$ $^{k}$ & $5.75$ $^{j}$   & $6.50$ $^{m}$ & $5.40$ $^{m}$ & $26$ $^{f}$         & $3.98$ $^{j}$ & $5.13$ $^{j}$      \\
$L[L_{\odot}]$ & $78$ $^{g}$          & $2430$ $^{k}$           & $9211$ $^{o}$          & $2512$ $^{m}$        & $2138$ $^{m}$        & $47039$ $^{h}$       & $224$ $^{g}$         & $155$ $^{o}$              \\
$\log$ g       & $4.22$ $^{j}$ & $4.06$ $^{p}$  & $3.84$ $^{p}$   & $3.62$ $^{m}$ & $3.77$ $^{m}$ & $3.27$ $^{p}$ & $3.75$ $^{j}$ & $3.49$ $^{j}$      \\
$[\mathrm{Fe/H}]$&$-0.08$ $^{p}$&$0.15$ $^{p}$  & $-0.07$ $^{p}$  & $0.25$ $^{i}$        & $-0.04$ $^{p}$& $-0.08$ $^{p}$& $0.06$ $^{p}$ & $- $                  \\

\hline\hline \noalign{\smallskip}
\end{tabular}
\end{center}
\tablefoot{
$^{(a)}$~\citet{2009A&A...501..297Z};  $^{(b)}$~\textmd{magnitudes from the General Catalogue of Photometric Data \citet{1997A&AS..124..349M}} ; $^{(c)}$~\citet{2007A&A...474..653V};  $^{(d)}$~\textmd{ derived from Eq.~\ref{equ.2} for stars with distances lower than 75 pc ($\lambda$ Aql, $\zeta$ Peg, and $\delta$ Cyg), and from Eq.~\ref{equ.2b} for more distant stars} $^{(e)}$~\textmd{average value based on the literature \citep{1994MNRAS.270..229W,2013ApJS..208....9P,2002yCat.2237....0D,1999PASP..111...63F}}, see the text for more explanations;  $^{(f)}$~\citet{2001A&A...367..521P}; $^{(g)}$~\citet{2012A&A...537A.120Z}; $^{(h)}$~\citet{2010AN....331..349H}; $^{(i)}$~\citet{1992ApJ...387..673G}; $^{(j)}$~\cite{1999A&A...352..555A}; $^{(k)}$~\cite{2013MNRAS.434.1321M}; $^{(l)}$~\cite{2011MNRAS.410..190T}; $^{(m)}$~\cite{2002MNRAS.333....9L}; $^{(n)}$~\citet{2005AJ....129.1642F}; $^{(o)}$~\cite{2012MNRAS.427..343M}; $^{(p)}$~\cite{2011A&A...525A..71W}.
}

\end{sidewaystable*}

This paper is structured as follows. Section~\ref{sect2} is devoted to a description of the stars in our sample. In Section~\ref{sect3}, we present the data reduction process and the method used to derive the angular diameters. Section~\ref{sect4} is dedicated to the calibration of the SBC relation, and we discuss our results in Section~\ref{sect5}.  We draw conclusions in Section~\ref{sect6}.

\section{VEGA/CHARA observations of eight early-type stars}\label{sect2}

We carefully selected eight early-type stars with a $(V-K)$ color index ranging from -0.70 to 0.02. They are north hemisphere main-sequence subgiant and giant stars ($\delta > 4^{o}$) with spectral types ranging from B1 to A1. They are much  brighter (with a visual magnitude $m_\mathrm{V}$ ranging from 1.6 to 4.7) than the limiting magnitude of VEGA (about $m_\mathrm{V}=7$ in medium spectral resolution). They are also bright in the K band (with a $m_\mathrm{K}$ magnitude lower than 5.1) which makes it possible to track the fringes simultaneously with the infrared CLIMB combiner \citep{2010SPIE.7734E.104S}. All the apparent magnitudes in V and K bands that we have collected from the literature are in the Johnson system (\citet{1966CoLPL...4...99J}, see also \citet{1997A&AS..124..349M}). The accuracy of their parallaxes $\pi$ spans from 1.5\% to 15\%. The color excess $E(V-K)$, the visual absorption $A_{V}$, the effective temperature $T_{\mathrm{eff}}$, the mass $M$, the radius $R$, the luminosity $L$, the surface gravity $\log g$, and the metallicity index $[\mathrm{Fe/H}]$ are listed in Table~\ref{Tab1}. We emphasize that for our purpose (limb-darkening estimates; see end of Sect.~\ref{sect3.1}), we do not need very precise estimates of the fundamental parameters of the stars in our sample, which explains why we do not provide any uncertainty on these parameters in the second part of Table~\ref{Tab1}.

Among the eight early-type stars in our sample, there are six low rotators ($\lambda$~Aql, $\gamma$~Ori, $\gamma$~Lyr, $\iota$~Her, 8~Cyg, and $\zeta$~Per) and two fast rotating stars ($\delta$~Cyg and $\zeta$~Peg). In the following, we define fast rotators as stars with $ v_\mathrm{rot} \sin i > 75 $~\kms. A theoretical study which aims at quantifying the impact of fast rotation on the SBC relation for early-type stars is currently in progress and will be published in a forthcoming paper.

We observed our sample stars from July 23, 2011, to August 29, 2013, using different suitable triplets available on the CHARA array. A summary of the observations is given in Table~\ref{Tab2}.

\begin{table}
\begin{center}
\caption[]{Summary of the observing log. All the details are given in the Appendix (Tables ~A\ref{Tabapx1},  A\ref{Tabapx2}, and A\ref{Tabapx3}). $N$ corresponds to the number of visibility measurements for each star. The reference stars used are also indicated (cf. Table~\ref{Tab3}).}
\label{Tab2}
\setlength{\doublerulesep}{\arrayrulewidth}
\begin{tabular}{lcccc}
\hline \noalign{\smallskip}
\hline\hline\hline\hline
  Name        &  3 telescope configurations & $N$  & Reference stars \\
\hline\hline
$\gamma$ Lyr  &    E2E1W2   &  23  & C6, C7        \\
$\gamma$ Ori  &  E2E1W2, W2W1S2, E2E1W2     &   8       & C1, C2, C3   \\
$8$ Cyg       &  W2W1E1   &  8  & C10         \\
$\iota$ Her   &  W2W1E1   &  8     & C4      \\
$\lambda$ Aql  &  S2S1W2   &  45    & C5, C9      \\
 $\zeta$ Per  &  E2E1W2   &  6    & C12       \\
$\delta$ Cyg  &   E2E1W2   &  22     & C8, C10     \\
  $\zeta$ Peg  &  E2E1W2   &  12    & C11       \\

    \hline\hline
\end{tabular}
\end{center}
\end{table}

\begin{table}
\begin{center}
\caption[]{Reference stars and their parameters, including the spectral type, the visual magnitude (m$_\mathrm{V}$), and the predicted uniform disk angular diameter (in mas) derived from the JMMC SearchCal software \citep{2006A&A...456..789B}. }
\label{Tab3}
\setlength{\doublerulesep}{\arrayrulewidth}
\begin{tabular}{p{0.4cm}p{1.4cm}p{0.9cm}p{0.3cm}c}
\hline \noalign{\smallskip}
\hline\hline\hline\hline
$\mathrm{No.}$ & Reference & S.Type      & m$_\mathrm{V}$  & $\theta_{UD}[R]$  \\
& stars & & & [mas] \\
\hline\hline
C1  & HD34989       & B1V      & 5.7	     &  0.130$\pm$0.009    \\
C2  & HD37320     & B8III    & 5.8	     &  0.153$\pm$0.011    \\
C3  & HD38899         & B9IV     & 4.8	     &  0.265$\pm$0.019    \\
C4  & HD167965           & B7IV     & 5.5      &  0.150$\pm$0.011 	\\
C5  & HD170296         & A1IV/V   & 4.6      &  0.429$\pm$0.031 	\\
C6  & HD174602           & A3V      & 5.2	     &  0.330$\pm$0.024  \\
C7  & HD178233           & F0III    & 5.5      &  0.399$\pm$0.029  \\
C8  & HD184875          & A2V      & 5.3      &  0.295$\pm$0.021    \\
C9 & HD184930          & B5III    & 4.3      &  0.317$\pm$0.022 	\\
C10 & HD185872    & B9III    & 5.4      &  0.200$\pm$0.014  \\
C11 & HD216735  & A1V      & 4.9	     &  0.310$\pm$0.022   	\\
C12 & HD22780  & B7Vn     & 5.5	     &  0.167$\pm$0.012  \\
\hline\hline
\end{tabular}
\end{center}
\end{table}

\begin{figure}[htbp]
 \begin{center}
\includegraphics[width=9cm]{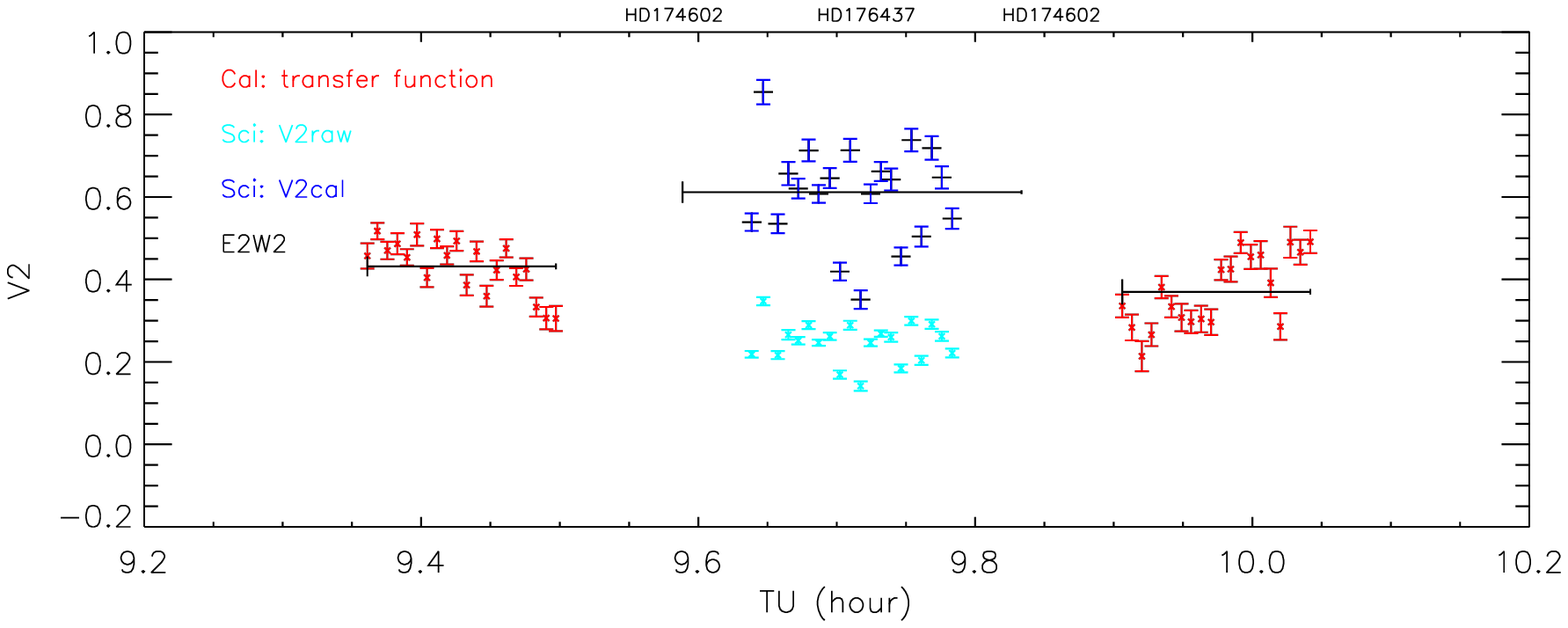}
\includegraphics[width=9cm]{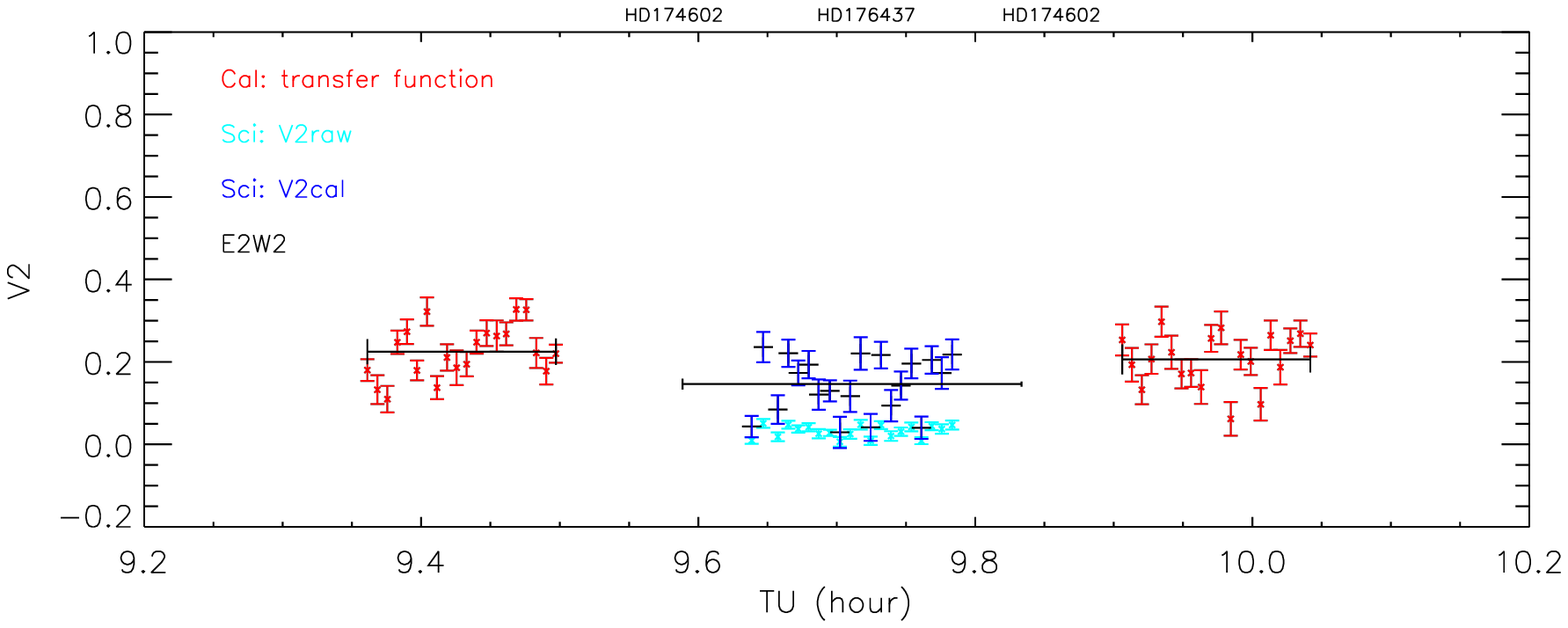}
\includegraphics[width=9cm]{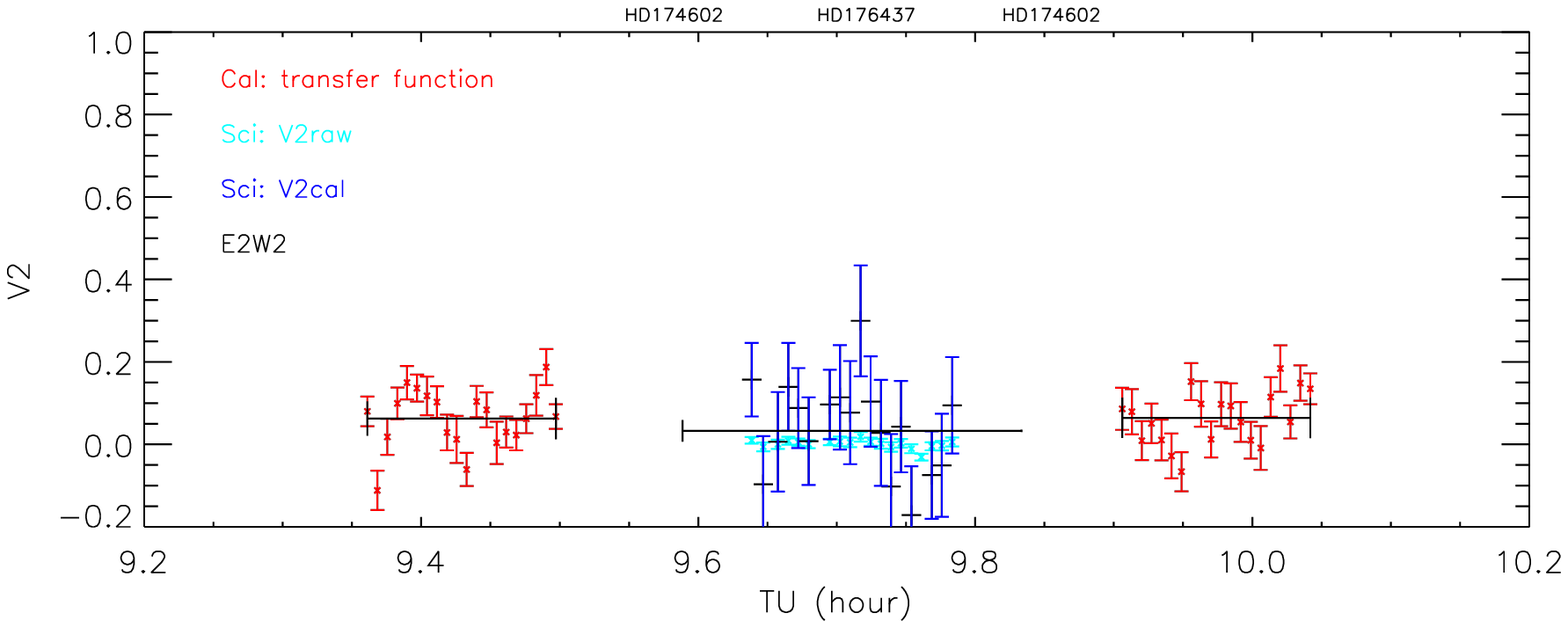}
\end{center}
\caption {The time sequences of raw visibilities of the science observations (light blue dots) are calibrated (blue dots) using the transfer function (red dots).}
  \label{fig1}
\end{figure}

\section{The limb-darkened angular diameters}\label{sect3}

In this section, we describe how we derive the limb-darkened angular diameter for all the stars in our sample.

\subsection{Data reduction and methodology}\label{sect3.1}

The first step is to calibrate the visibility measurements of our targets using observations of reference stars. These calibrators (Table~\ref{Tab3}) were selected using the SearchCal\footnote[1]{Available at \url{http://www.jmmc.fr/searchcal}} software developed by the Jean-Marie Mariotti Center (JMMC) \citep{2006A&A...456..789B}. The way this calibration is done is shown in Fig.~\ref{fig1} in the case of $\gamma$~Lyr (data obtained on June 21, 2012, with the E1E2W2 three-telescope configuration). We used the standard sequence C6-S-C6 in which S is the target and C6 is the reference star.  The light blue dots are the raw visibilities obtained on the science star for the three corresponding baselines: E2E1 (upper panel), E2W2 (middle panel), and E1W2 (bottom panel). Our VEGA measurements are typically divided into 30 blocks of observations, and each block contains 1000 images with an exposure time of 15 milliseconds. For each block, the raw squared visibility is calculated using the auto-correlation mode \citep{2009A&A...508.1073M, 2011A&A...531A.110M}. The red dots in the figure represent the transfer function obtained by comparing the expected visibility of the reference star to the one that has been measured. This transfer function is then used to calibrate the visibilities obtained on the science target (blue dots). A cross-check of the quality of the transfer function is usually done for several bandwidths and over the whole night. Under good seeing conditions, the transfer function of VEGA/CHARA is generally stable at the level of 2\% for more than one hour. The squared calibrated visibilities $V^{2}_{\mathrm{target}}$ obtained from our VEGA observations are listed in Tables ~A\ref{Tabapx1},  A\ref{Tabapx2}, and A\ref{Tabapx3}. The systematic uncertainties that stem from the uncertainty on the reference stars are negligible compared to the statistical uncertainties, and are neglected in the rest of this study.

\begin{table*}
\begin{center}
\caption[]{Angular diameters obtained with VEGA/CHARA and the corresponding surface brightness. The systematical uncertainties for the two fast rotating stars, $\zeta$~Peg  and $\delta$~Cyg, are of $0.039$~mas and $0.047$~mas, respectively (see section~\ref{sect3.2}).}
\label{Tab5}
\setlength{\doublerulesep}{\arrayrulewidth}
\begin{tabular}{lcccccc}
\hline \noalign{\smallskip} \hline\hline\hline\hline
Star & $(V-K)_{0}$ & $\theta_{\mathrm{UD}} [mas]$ & $\chi^{2}$ & $U_{R}$ & $\theta_{\mathrm{LD}}$ [mas]& $S_{\mathrm{v}}[mag]$ \\
\hline\hline
$\lambda$ Aql & $-0.265\pm0.055$ & $0.529\pm0.003$ & $1.0$ & $0.301$  &$0.544\pm0.003$ & $2.079\pm0.030$ \\
$\gamma$ Lyr  & $-0.102\pm0.072$ & $0.742\pm0.010$ & $2.9$ & $0.402$  &$0.766\pm0.010$ & $2.544\pm0.059$ \\
$\gamma$ Ori  & $-0.703\pm0.097$ & $0.701\pm0.005$ & $0.4$ & $0.269$  &$0.715\pm0.005$ & $0.909\pm0.081$ \\
$8$ Cyg       & $-0.492\pm0.147$ & $0.229\pm0.011$ & $1.3$ & $0.299$  &$0.234\pm0.011$ & $1.456\pm0.177$ \\
$\iota$ Her   & $-0.459\pm0.076$ & $0.304\pm0.010$ & $1.2$ & $0.280$  &$0.310\pm0.010$ & $1.225\pm0.082$ \\
$\zeta$ Per   & $-0.592\pm0.092$ & $0.531\pm0.007$ & $1.2$ & $0.343$  &$0.542\pm0.007$ & $0.652\pm0.081$ \\
$\zeta$ Peg   & $-0.204\pm0.055$ & $0.539\pm0.009$ & $1.7$ & $0.442$  &$0.555\pm0.009$ & $2.076\pm0.152$ \\
$\delta$ Cyg  & $+0.021\pm0.055$ & $0.766\pm0.004$ & $1.3$ & $0.408$  &$0.791\pm0.004$ & $2.318\pm0.129$ \\

\hline\hline
\end{tabular}
\end{center}
\end{table*}

\begin{figure*}
 \begin{center}
\includegraphics[width=8.25cm]{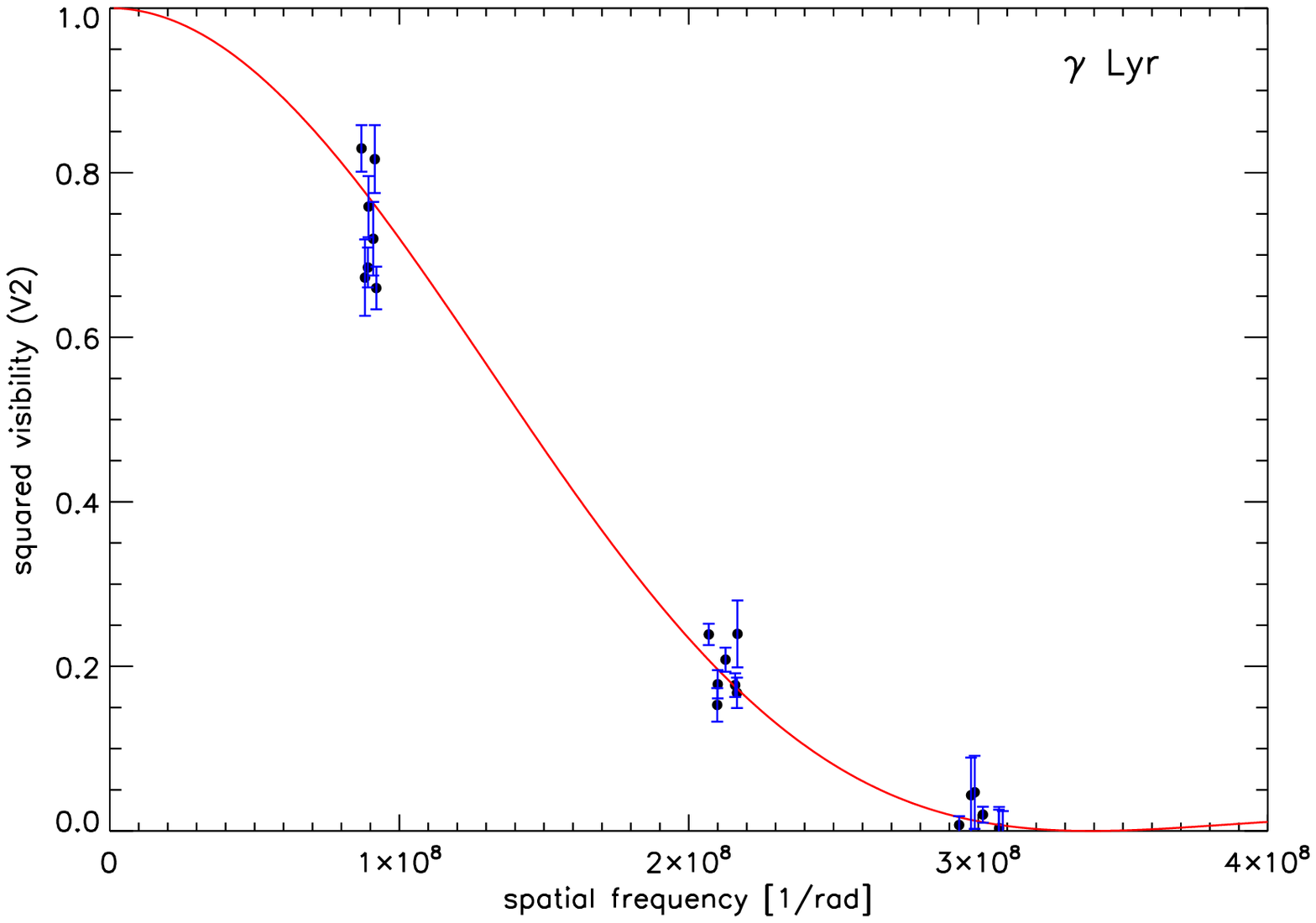}
\includegraphics[width=8.25cm]{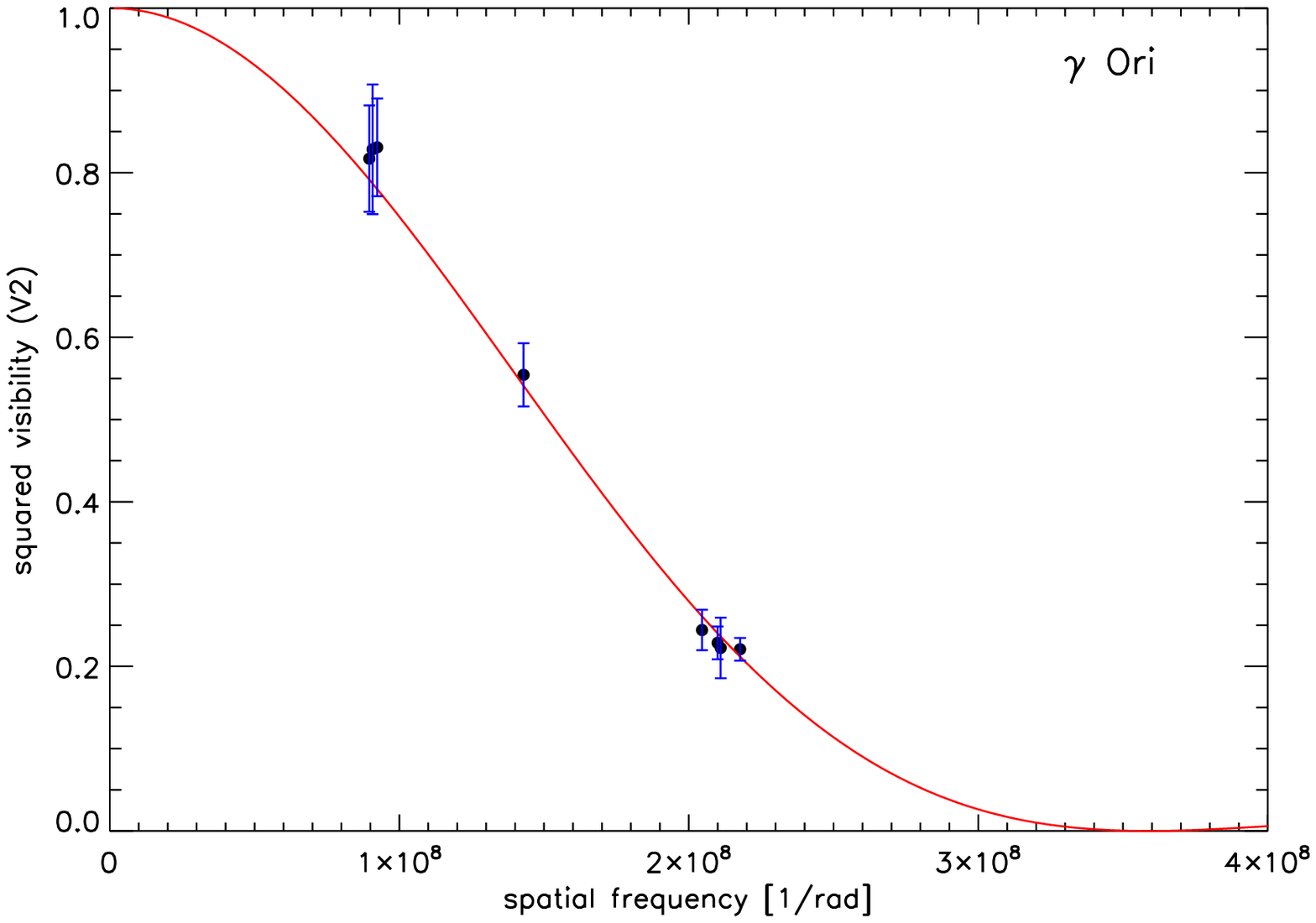}
\includegraphics[width=8.25cm]{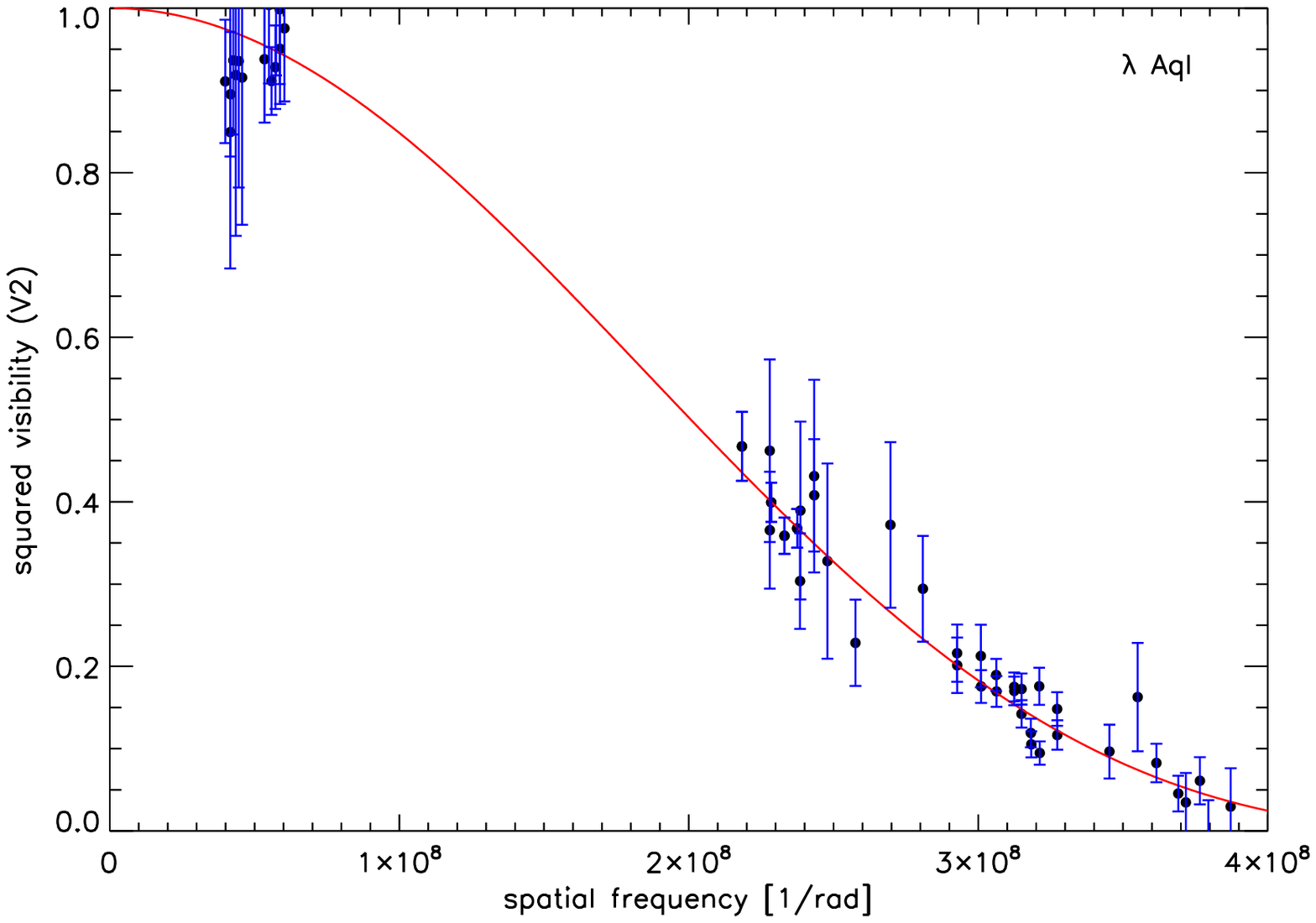}
\includegraphics[width=8.25cm]{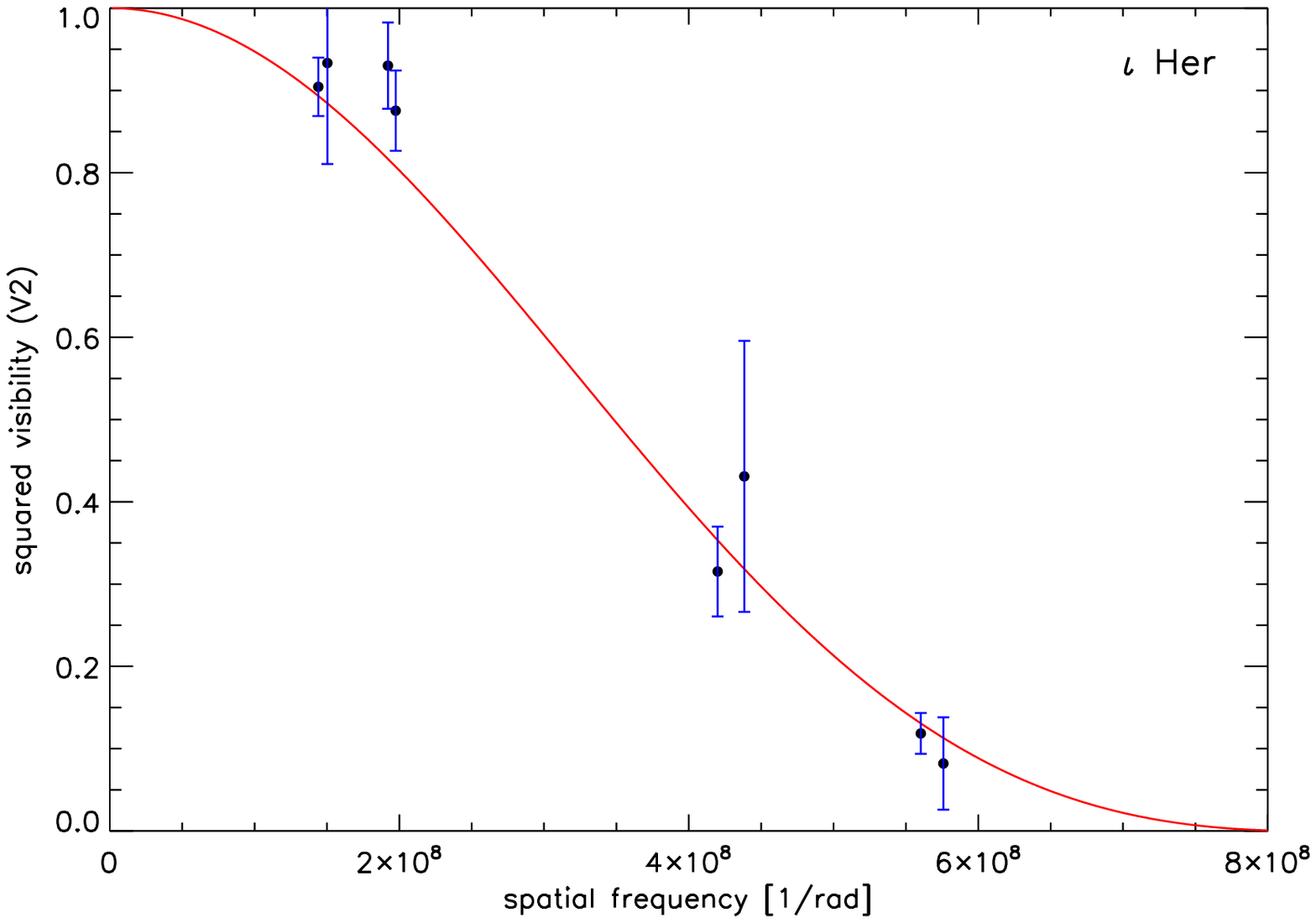}
\includegraphics[width=8.25cm]{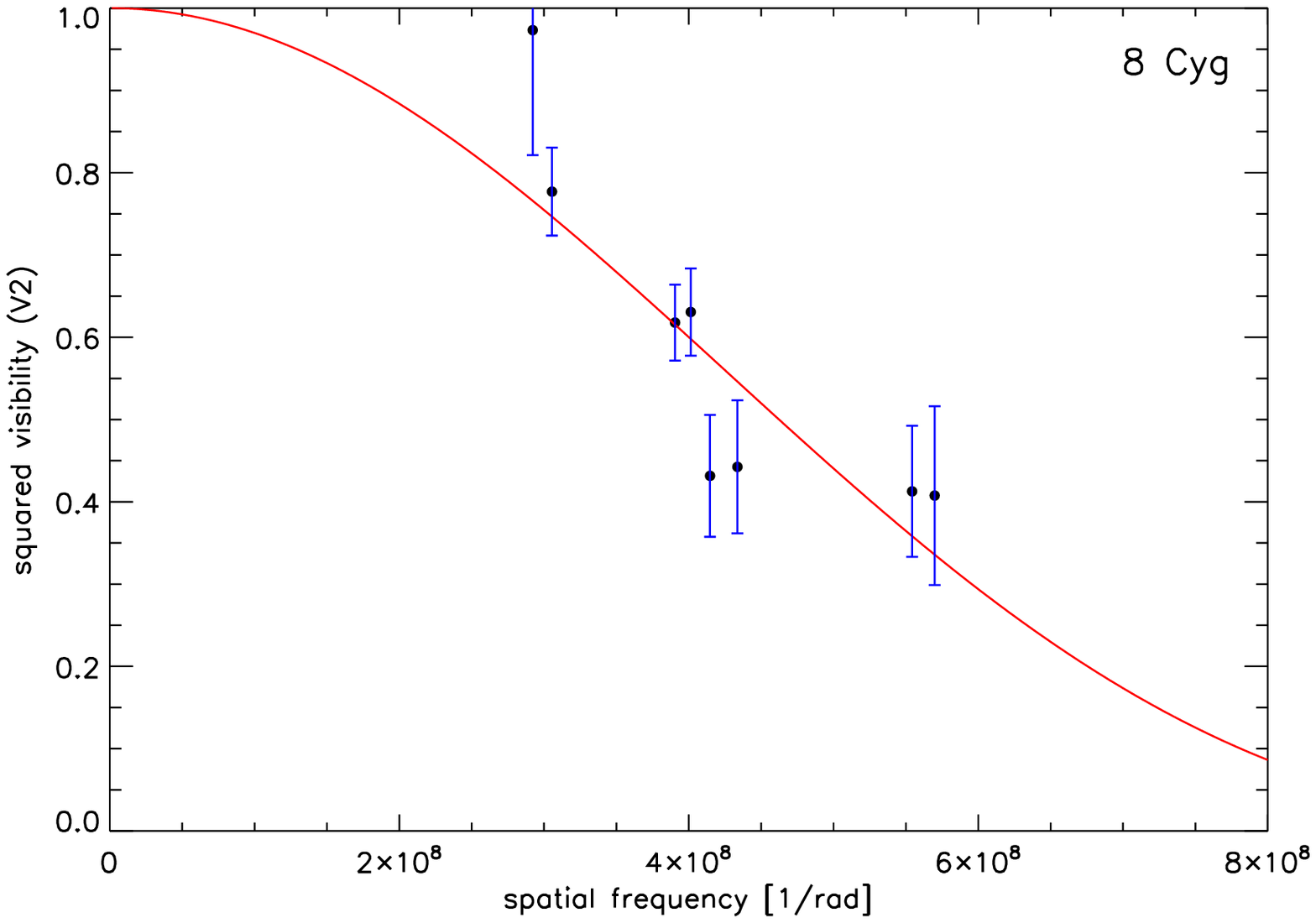}
\includegraphics[width=8.25cm]{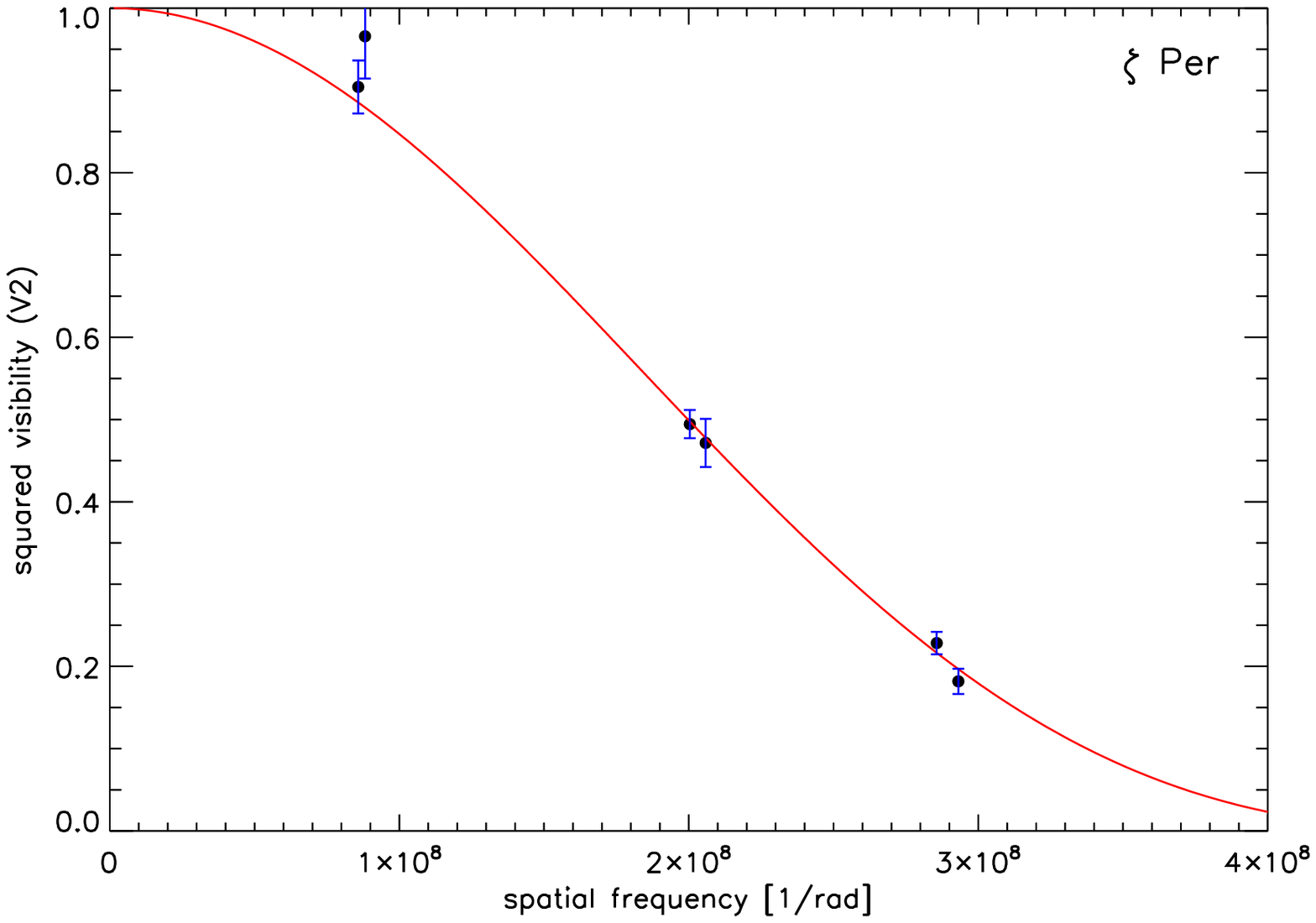}
\includegraphics[width=8.25cm]{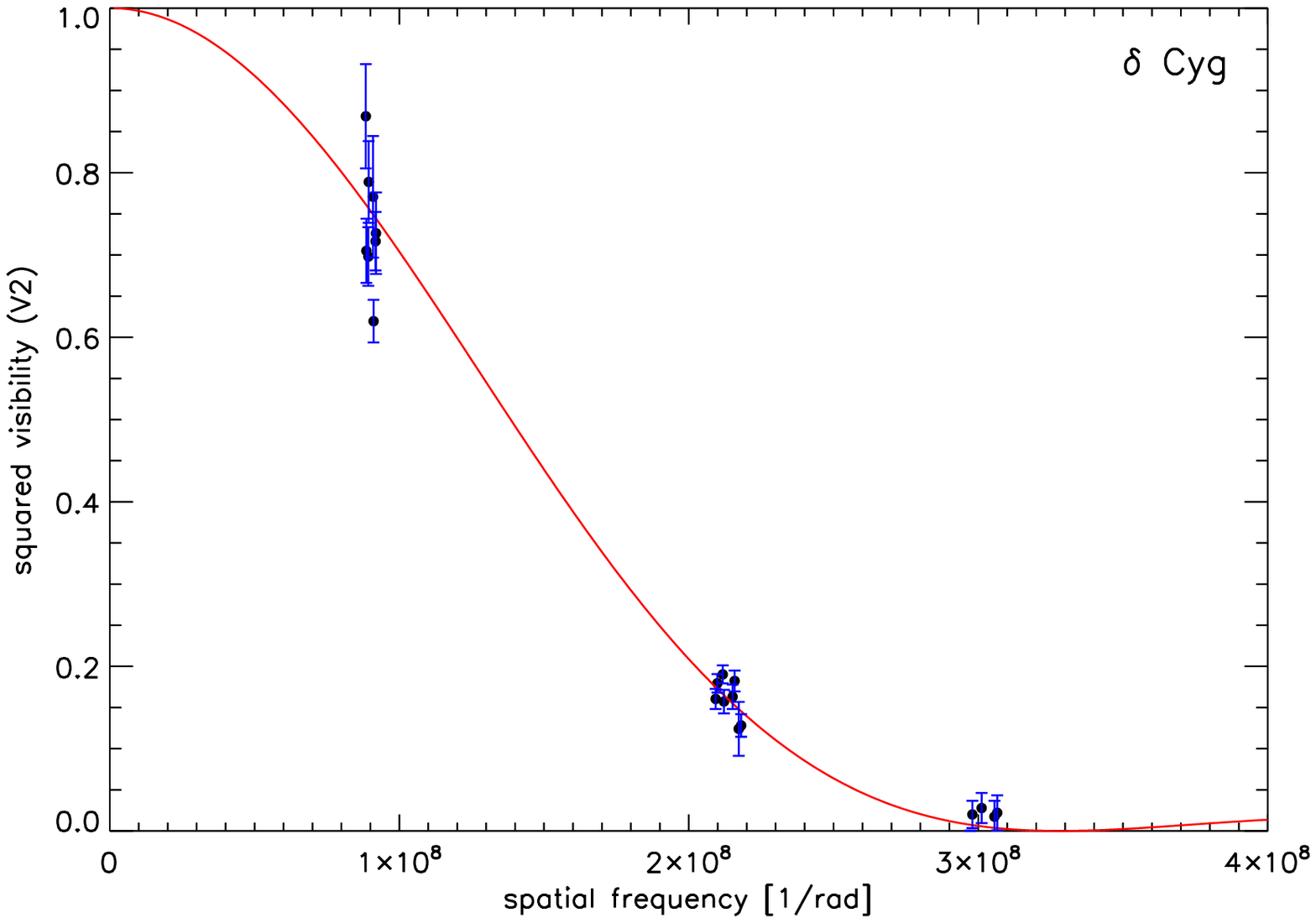}
\includegraphics[width=8.25cm]{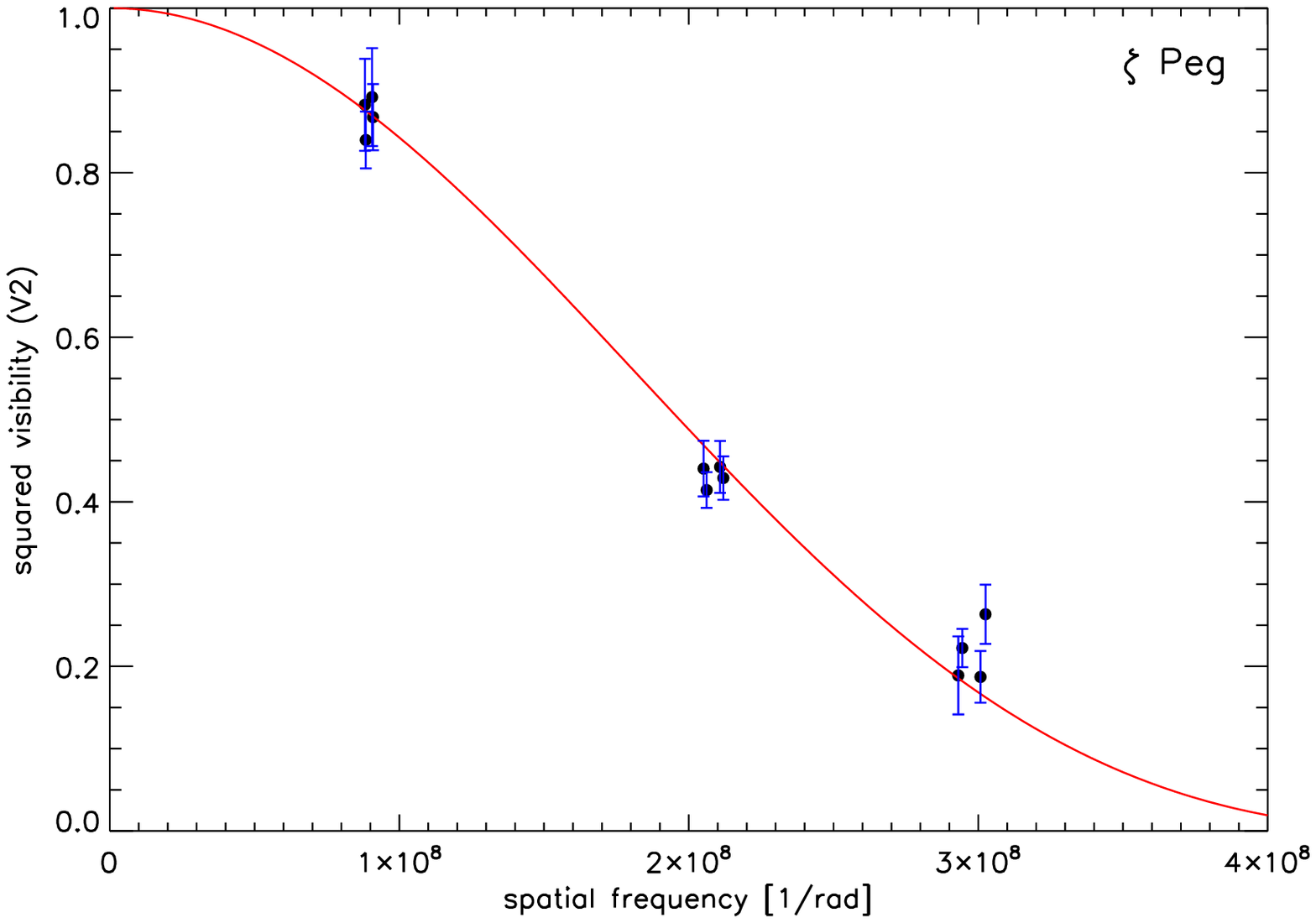}
\caption {Squared visibility versus spatial frequency for all stars in our sample with their corresponding statistical uncertainties. The red solid lines indicate the best uniform disk model obtained from the LITpro fitting software.} \label{fig22}
\end{center}
\end{figure*}

The calibrated visibility curves obtained for each star in our sample (Fig.~\ref{fig22}) are then used to constrain a model of uniform disk, that contains only one parameter, the so-called uniform disk angular diameter ($\theta_{\mathrm{UD}}$).  This is performed using the LITpro\footnote[2]{Available at \url{http://www.jmmc.fr/litpro}} software developed by the JMMC \citep{2008SPIE.7013E..44T}. The following formula of \citet{1974MNRAS.167..475H} provides an analytical way to convert the equivalent uniform disk angular diameter $\theta_{\mathrm{UD}}$ into the limb-darkened disk $\theta_{\mathrm{LD}}$:

\begin{equation}
\theta_{\mathrm{LD}}(\lambda)=\theta_{\mathrm{UD}}(\lambda)\left[
                 \frac{(1-\frac{U_{\lambda}}{3})}{(1-\frac{7U_{\lambda}}{15})}
              \right]
              ^{\frac{1}{2}}\cdot
\end{equation}

For each star, the limb-darkening coefficient $U_{\lambda}$ is derived from the numerical tables of \citet{2011A&A...529A..75C}. These tables are based on the ATLAS \citep{1970SAOSR.309.....K} and PHOENIX \citep{1997ApJ...483..390H} atmosphere models. The input parameters of these tables are the effective temperature ($T_\mathrm{eff}$),  the metallicity ($[\mathrm{Fe/H}]$), the surface gravity ($\log g$), and the micro-turbulence velocity. The steps for these quantities are 250K, 0.5, 0.5, and 2 \kms, respectively. The three first parameters are given in Table~\ref{Tab1} and were rounded, for each star,  to the closest value found in the table of  Claret. The micro-turbulence velocity has almost no impact on the derived limb-darkened diameter (fifth decimal). We took arbitrarily 8~\kms for stars with $T_{\mathrm{eft}}>15000K$ and 4~\kms for stars with $T_{\mathrm{eft}}<15000K$. We also consider the limb-darkening coefficient applicable to the $R$ band of VEGA ($U_\mathrm{R}$ in the following).

\subsection{Results}\label{sect3.2}

The uniform disk angular diameter ($\theta_{\mathrm{UD}}$), the limb-darkening coefficients ($U_{\mathrm{R}}$), and the derived limb-darkened angular diameters ($\theta_{\mathrm{LD}}$) are listed in Table~\ref{Tab5} for each star in our sample. The value of $\theta_{\mathrm{LD}}$ ranges from 0.31 mas to 0.79 mas, with a relative precision from 0.5\% to 3.5\% (average of 1.5\%). The reduced $\chi_{\mathrm{red}}^2$ is from 0.4 to 2.9 depending on the dispersion of the calibrated visibilities. For  $\gamma$~Lyr, our result ($\theta_{\mathrm{UD}}=0.742\pm0.010$~mas) agrees at the 1$\sigma$ level with the measurements from the PAVO/CHARA instrument ($\theta_{\mathrm{UD}}=0.729 \pm 0.008$~mas, \citet{ 2013MNRAS.434.1321M}). For $\gamma$~Ori, our angular diameter ($\theta_{\mathrm{LD}}=0.715\pm0.005$~mas) is consistent with the value derived from the Narrabri Stellar Intensity Interferometer (NSII) ($\theta_{\mathrm{LD}}=0.72\pm0.04$~mas, \cite{1974MNRAS.167..121H}). For other stars with angular diameters lower than 0.6~mas ($\iota$~Her, $\lambda$~Aql, 8~Cyg, and $\zeta$~Per) and for the two fast rotators ($\zeta$~Peg and $\delta$~Cyg) there are no interferometric observations available to our knowledge.

 For the two rotators, we derive the apparent oblateness using the approximate relation provided by \cite{2006ApJ...637..494V}, their Eq.~A1 $ \frac{R_{\mathrm{b}}}{R_{\mathrm{a}}} \simeq 1-(v \sin i)^2  \frac{R_{\mathrm{b}}}{2 G M}$, where $R_{\mathrm{b}}$, $R_{\mathrm{a}}$, M, and G are the major and minor apparent radius of the star, its mass, and the gravitational constant. We find  $ \frac{R_{\mathrm{b}}}{R_{\mathrm{a}}} = 1.07$ for  $\zeta$~Peg considering $R_{\mathrm{b}} \simeq \overline{R} = 4.03\Ro$ and $M=3.22\Mo$, where  $\overline{R}$ is the mean radius (see Table~\ref{Tab1}), while the rotational projected velocity $v \sin i$ is set to 140~\kms  \citep{2002ApJ...573..359A}. For $\delta$~Cyg, we find similarly $ \frac{R_{\mathrm{b}}}{R_{\mathrm{a}}} = 1.06$ considering $v \sin i = 140$~\kms \citep{1975ApJS...29..137S, 1980PASP...92..771G, 1984ApJ...286..741C, 1995ApJS...99..135A, 2002ApJ...573..359A, 2012A&ARv..20...51V}. Consequently, our data might be sensitive to the expected gravity darkening intensity distribution and the flatness of the star. However, this also depends on the baseline orientation. For both stars, the three telescopes (Table~\ref{Tab2}) are aligned. Thus, even if our reduced $\chi_{\mathrm{red}}^2$ are rather low (1.7 for $\zeta$~Peg  and 1.2 for $\delta$~Cyg), we cannot  exclude a bias on our derived limb-darkened angular diameters. In order to get a rough estimate of this bias, we only consider in first approximation the oblateness of the star while the gravity darkening is set to be negligible. As a consequence, if the orientation of the baseline is aligned with the polar or equatorial axis, we can estimate a maximum systematic error of about $0.039$~mas (6\%) for $\zeta$~Peg, while we find $0.047$~mas (7\%) for $\delta$~Cyg.  We translate these uncertainties in terms of $S_{\mathrm{v}}$ magnitude in Sect.~\ref{sect4}.

\section{The calibration of the surface brightness relation}\label{sect4}

\subsection{Methodology}\label{sect4.11}

As already mentioned in the introduction, the SBC relation is a very robust tool for the distance scale calibration. The surface brightness S$_{V}$ of a star is linked to its visual intrinsic dereddened magnitude $m_\mathrm{V_{\mathrm{0}}}$ and its limb-darkened angular diameter $\theta_{\rm LD}$ by the following relation:

\begin{equation}\label{equ.1}
S_{\mathrm{V}}=m_\mathrm{V_{\mathrm{0}}}+5\log \theta_{LD}\cdot
\end{equation}

Instead of S$_{V}$, the surface brightness parameter $F_{\mathrm{V}}=4.2207 - 0.1 S_{\mathrm{V}}$ is often adopted in the literature to determine the stellar angular diameters \citep{1976MNRAS.174..489B}. In order to derive $m_\mathrm{V_{\mathrm{0}}}$, we first selected the apparent $m_\mathrm{V}$ magnitudes for all the stars in our sample \citep{1997A&AS..124..349M}. These magnitudes are expressed in the Johnson system \citep{1966CoLPL...4...99J} and their typical uncertainty is of about $0.015$ mag. In order to correct these magnitudes from the reddening  we then use the following formulae $ m_\mathrm{V_{\mathrm{0}}}=m_\mathrm{V}-A_{V}$, where $A_{V}$ is the extinction in the V band. Determining the extinction is a difficult task. We adopt the following strategy. For stars lying closer than 75 pc we use the simple relation

\begin{equation}
A_{V} = \frac{0.8}{\pi},
\label{equ.2}
\end{equation}
where $\pi$ is the parallax of the stars [in mas]. This equation is standard in the literature \citep{1990A&A...232..396B, 1998A&A...339..858D, 2005MNRAS.357..174D}.
The corresponding uncertainty is set to 0.01 magnitude.

For distant stars we derive the absorption using the (B-V) extinction \citep{1993MNRAS.263..921L}:

\begin{equation}
A_{V} = 3.1E(B-V)\cdot
\label{equ.2b}
\end{equation}

The difficulty is then to derive $E(B-V)$. We have several possibilities. First, we use the so-called Q method, with $Q=(U-B)-0.72(B-V)$, which was originally proposed by \citet{1953ApJ...117..313J}. The value of Q is derived for each star using observed UBV magnitudes from \citep{2002yCat.2237....0D}. Then a relation between $(B-V)_{\mathrm{0}}$ and Q can be found in \citet{2013ApJS..208....9P}, and $E(B-V)$ is finally derived using $E(B-V)=(B-V)-(B-V)_{\mathrm{0}}$.

Second, from the spectral type of the stars in our sample, we can derive their intrinsic colors in different bands using Table~5 of \citet{1994MNRAS.270..229W}. We thus obtain $(B-V)_{\mathrm{0}}$, $(V-R)_{\mathrm{0}}$, $(V-I)_{\mathrm{0}}$, $(V-J)_{\mathrm{0}}$, $(V-H)_{\mathrm{0}}$, and $(V-K)_{\mathrm{0}}$. Once compared with the observed colors from \citet{2002yCat.2237....0D}, we derive E(B-V), E(V-R), E(V-I), E(V-J), E(V-H), and E(V-K), and we finally use Table~2 (col. 4) from \citet{1999PASP..111...63F} and assume total to selective extinction ratio in B-band $\frac{A_{B}}{E(B-V)}=4.1447$ (Table 3 from \citet{1989ApJ...345..245C}) to perform a conversion into E(B-V) using the following equations:

\begin{equation}\label{equ.11}
E(B-V)=\frac{E(V-R)(A_{B}-A_{V})}{(A_{V}-A_{R})}=1.2820 E(V-R)
\end{equation}
\begin{equation}\label{equ.12}
E(B-V)=\frac{E(V-I)(A_{B}-A_{V})}{(A_{V}-A_{I})}=0.6536 E(V-I)
\end{equation}
\begin{equation}\label{equ.13}
E(B-V)=\frac{E(V-J)(A_{B}-A_{V})}{(A_{V}-A_{J})}=0.4464 E(V-J)
\end{equation}
\begin{equation}\label{equ.14}
E(B-V)=\frac{E(V-H)(A_{B}-A_{V})}{(A_{V}-A_{H})}=0.3891 E(V-H)
\end{equation}
\begin{equation}\label{equ.15}
E(B-V)=\frac{E(V-K)(A_{B}-A_{V})}{(A_{V}-A_{K})}=0.3650 E(V-K)\cdot
\end{equation}

We finally obtain seven values of the extinction (Q method, and six values derived from Table~5 of \citet{1994MNRAS.270..229W}). These quantities are averaged and their statistical dispersion provides a realistic uncertainty (indicated in Table~\ref{Tab1} for the VEGA sample). However, the Q method is applicable only for stars of class IV and V, while Table~5  of \citet{1994MNRAS.270..229W} can be used only for spectral types O and B. We thus have in some cases fewer than seven values. And even, in the case of  $\gamma$~Lyr, for instance (which is an A1III star standing at a distance greater than 75 pc), we used other $E(B-V)$ estimates available in the literature (see Table~\ref{Tab1}). The uncertainty on $S_{\mathrm{V}}$ is finally derived from the uncertainty on $m_\mathrm{V}$ (typically 0.015), the angular diameter (see Table~\ref{Tab5}), and $A_{V}$.

In order to mitigate the effects from a somewhat erroneous calibration of the intrinsic colors, we recalculate $(V-K)_\mathrm{0}$ from the derived E(B-V) value. First we calculate E(B-V) from averaging via Eq.~\ref{equ.11}-~\ref{equ.15}. Then using this value and Eq.~\ref{equ.15}, we derive E(V-K) (given in Table~\ref{Tab1}). From E(V-K), $m_\mathrm{V}$, and $m_\mathrm{K}$  we obtain $(V-K)_\mathrm{0}$.  The uncertainty on $(V-K)_{\mathrm{0}}$ is derived assuming an uncertainty of $0.015$ for $m_{\mathrm{V}}$, 0.03 for $m_{\mathrm{K}}$ (following  \citet{2005MNRAS.357..174D}), and the uncertainty on E(V-K), itself derived from the uncertainty obtained on E(B-V). The $m_\mathrm{V}$ and $m_\mathrm{K}$ magnitudes for the stars in our sample are given in Table~\ref{Tab1} together with $\pi$, $E(V-K)$, and $A_{V}$. The derived values of the surface brightness for each star are given in Table~\ref{Tab5}.

In order to calibrate the SBC relation,  we also need to combine the eight limb-darkened angular diameters derived from the VEGA observations with different  sets of diameters already available in the literature.

\begin{figure}[htbp]
\begin{center}
\includegraphics[width=9cm,height=7cm]{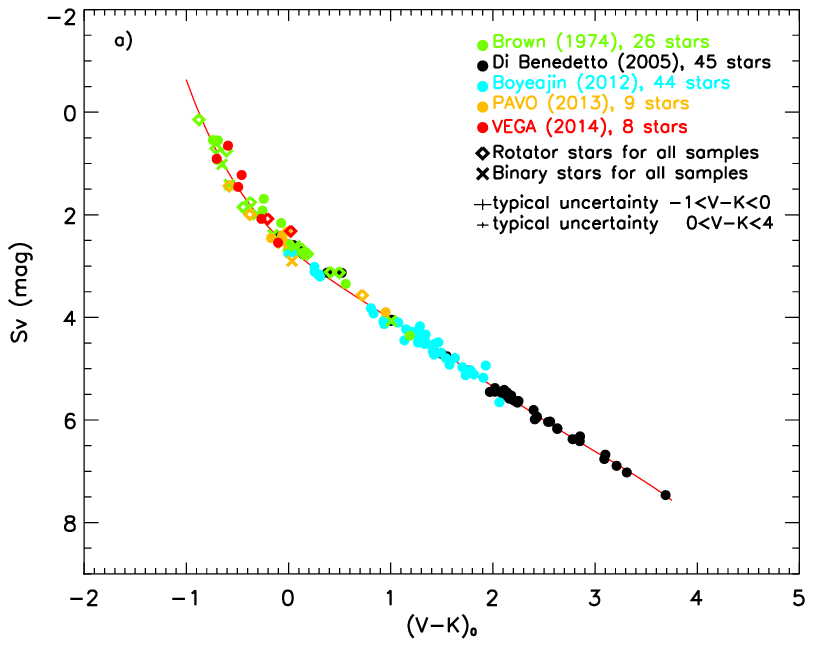}
\includegraphics[width=9cm,height=3cm]{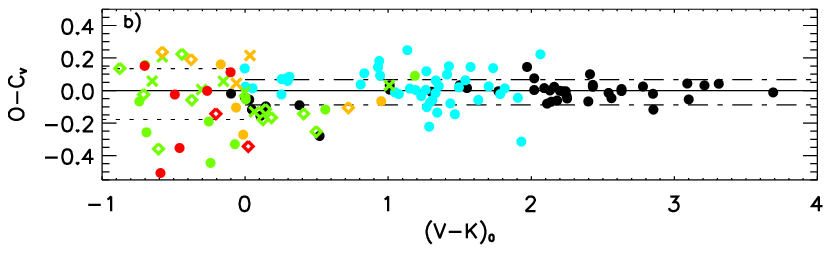}
\includegraphics[width=9cm,height=3cm]{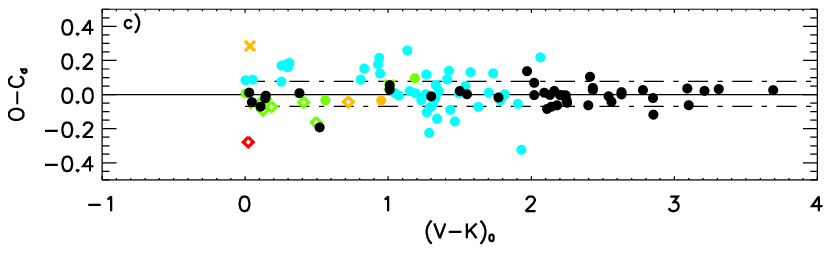}
\includegraphics[width=9cm,height=3cm]{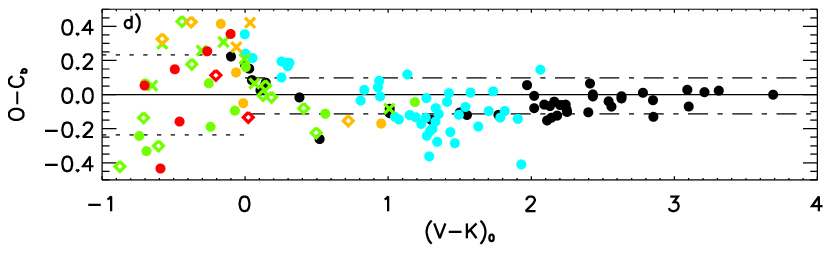}
\end{center}
\caption { The relation between visual surface brightness $S_{V}$ as a function of the color index $(V-K)_{0}$. The black, light blue, green, brown, and red measurements are from \citet{2005MNRAS.357..174D},  \citet{2012ApJ...746..101B},  \cite{1974MNRAS.167..121H}, \cite{2013MNRAS.434.1321M}, and VEGA (this work), respectively. The red line corresponds to our fit when considering all stars. The rms of the difference between the surface brightness computed from our fit and measured surface brightness is presented in the lower panels (see the text for more detail).}
  \label{fig3}
\end{figure}

\begin{figure}[htbp]
 \begin{center}
 \includegraphics[width=9cm,height=7cm]{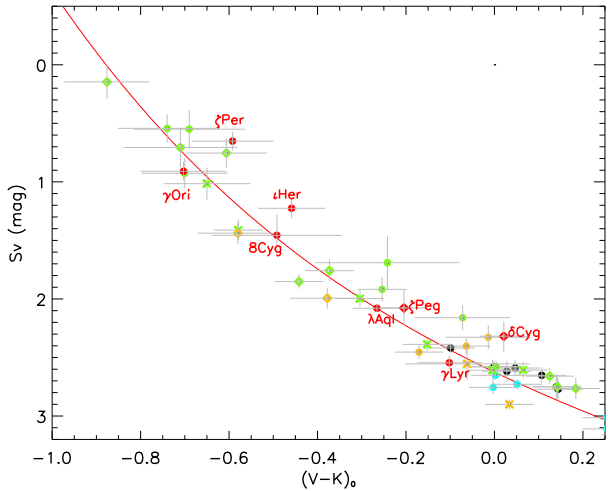}
\end{center}
\caption {Same as Fig.\ref{fig3}, but with the names of the VEGA/CHARA stars in our sample (in red).}
  \label{fig4}
\end{figure}

\subsection{A revised SBC relation for late- and early-type stars}\label{sect.4.1}

Historically, the SBC was first derived from interferometric observations of 18 stars by \citet{1969MNRAS.144..297W} using the (B-V) index. Five years later, the apparent angular diameters of 32 stars in the spectral range O5 to F8 have been measured using the NSII \citep{1974MNRAS.167..121H}. Based on this sample, \citet{1976MNRAS.174..503B}  and \citet{1976MNRAS.174..489B} calibrated the SBC for late-type and early-type stars, respectively, but this was not done with the V-K color index. In order to constrain the SBC relation as a function of V-K we therefore use these 32 angular diameters (but 6 are rejected, see below). This is the first set of data we have used. We emphasize that the $(V-K)_{0}$ color index is usually used to calibrated the SBC relation because it provides the lowest rms and it is mostly parallel to the reddening vector on the Sv - (V - K) diagram. Moreover, for all the datasets we have considered, we have recalculated the $(V-K)_{0}$ and $A_{V}$ values in a similar way as for the VEGA objects (see Sect.~\ref{sect4.11}).

More than ten years later, \citet{1998A&A...339..858D} made a careful compilation of 22 stars (with A, F, G, K spectral types) for which angular diameters were available in the literature and calibrated the SBC relation. Moreover, the direct application of the SBC relation to Cepheids was done by \citet{1997A&A...320..799F} and \citet{1998A&A...339..858D}. Later, 27 stars were measured by NPOI and Mark III optical interferometers and the derived high precision angular diameters were published by \citet{2001AJ....122.2707N} and \citet{2003AJ....126.2502M}, respectively. Finally, using a compilation of 29 dwarfs and subgiant (including the sun) in the $0.0\leq(V-K)_{0}\leq6.0$ color range, \citet{2004A&A...426..297K} calibrated for the first time a linear SBC relation with an intrinsic dispersion of 0.02 mag or 1\% in terms of angular diameter. A short time later, \citet{2005MNRAS.357..174D} made the same kind of compilation but with 45 stars in the $-0.1\leq(V-K)_{0}\leq3.7$ color range (accuracy of 0.04 mag or 2\% in terms of angular diameter). We use this larger second data set for our analysis.

One year later, \citet{2006A&A...456..789B} provided a SBC relation (as function of $V-K$ color magnitude) based on interferometric measurements, lunar occultation, and eclipsing binaries.  We compare our results with those of \citet{2006A&A...456..789B} and also \citet{2005MNRAS.357..174D} in Sect.~\ref{sect5}.

Recently, \citet{2012ApJ...746..101B} enlarged the sample to 44 main-sequence A-, F-, and G-type stars using CHARA array measurements. In addition, ten stars with spectral types from B2 to F6 were observed using the Astronomical Visible Observations (PAVO) beam combiner at the CHARA array \citep{2013MNRAS.434.1321M} and these recent CHARA measurements have been incorporated in our analysis.

However, in order to derive a SBC relation accurate enough for distance determination, one has to perform a consistent selection. Our strategy is the following: we consider all stars in multiple systems (as soon as the companion is far and faint enough not to contaminate interferometric measurements), fast rotators, and single stars. Fast rotating stars should be included as they improve the statistics of the relation (in particular for early-type objects), even if a slight bias is not excluded as discussed in Sect.~\ref{sect3.2} (see also next section). Conversely, we exclude stars with environments (like Be stars with strong wind) or stars with a strong variability. Following these criteria, we found seven stars to reject. The first one is Zeta Orionis ($\zeta$~Ori). Its angular diameter, measured by \citet{1974MNRAS.167..121H} is most probably biased by a companion which was discovered later and with a separation of 40~mas \citep{2000ApJ...540L..91H,2013A&A...554A..52H}. Kappa Orionis ($\kappa$~Ori) shows a P~Cygni profile in H$\alpha$ caused by a stellar wind \citep{2008A&A...481..777S,1981A&A...101..168S, 1983ApJ...268..205C}. Delta Scorpii ($\delta$~Sco) is an active binary star exhibiting the Be phenomenon \citep{2013A&A...550L...5M}. Gamma2 Velorum ($\gamma^2$~Vel) is a binary system with a large spectral contribution from the Wolf-Rayet star \citep{2007A&A...464..107M}. Zeta Ophiuchi ($\zeta$~Oph) is a magnetic star of $O_{\mathrm{e}}$-type \citep{2011AN....332..147H}. Alpha Virginis ($\alpha$~Vir) is a double-lined spectroscopic binary (B1V+B4V) with an ellipsoidal variation of 0.03~mag due to tidal distortion \citep{2009ApJ...704..813H}. The last one, Zeta Cassiopeiae ($\zeta$~Cas), is in the PAVO sample \citep{2013MNRAS.434.1321M}. It stands at 7$\sigma$ from the relation. It is a $\beta$~Cepheid and the photometric contamination by a surrounding environment and/or a close companion is not excluded \citep{1994A&A...287..509S, 2011A&A...525A..67N}.

We finally end with 26 stars from \citet{1974MNRAS.167..121H}, 44 stars from \citet{2012ApJ...746..101B}, 9 stars from \citet{2013MNRAS.434.1321M}, and 45 values of $S_{v}$ from \citet{2005MNRAS.357..174D}, to which we can add our eight angular diameters obtained with VEGA/CHARA. The total sample is composed of 132 stars (with $-0.876<V-K<3.69$), including 32 early-type stars with -1<V-K<0. Using this sample of 132 stars, we find the relation

\begin{equation}\label{final}
S_{v}=\sum_{n=0}^{n=5}C_{n}(V-K)^{n}_{0}
\end{equation}
with, $C_{0}=2.624\pm0.009$,    $C_{1}=1.798\pm0.020$,    $C_{2}= -0.776\pm0.034$,   $C_{3}= 0.517\pm0.036$,   $C_{4}=-0.150\pm0.015$, and $C_{5}=0.015\pm0.002$. Uncertainties on coefficients of the SBC relation do not take into account the X-axis uncertainties on $(V-K)_\mathrm{0}$. This relation can be used consistently in the range $-0.9 \leq V-K \leq 3.7$ with $\sigma_\mathrm{S_{v}}$=0.10 mag. This corresponds to a relative precision on the angular diameter of $ \frac{\sigma \theta}{\theta}  = 46.1 \sigma_{\mathrm{S_\mathrm{V}}} \simeq 4.6 \% $ derived from Eq.~5 of \citet{2005MNRAS.357..174D}. For stars earlier than A3 ($-0.9 < V-K < 0.0$), we successfully reached a magnitude precision of $\sigma=0.16$ or 7.3\% in terms of angular diameter.

\section{Discussion}\label{sect5}

Figure~\ref{fig3}a shows the resulting SB relation as a function of the $(V-K)_\mathrm{0}$ color index for the five different data sets we have considered. The VEGA data appear in red in the figure. The residual O-C$_{v}$, which is the difference obtained between the measured surface brightness (O) and the relation provided by Eq.~\ref{final} (C$_{v}$), is shown in Fig.~\ref{fig3}b. In the following, we define $\sigma_{+}$ and $\sigma_{-}$ as the positive and negative standard deviation. We obtain $\sigma_{+}=0.07$ and  $\sigma_{-}=0.09$ for $0 < V-K <3.7$ (late-type stars, dot-dashed line in the figure) and $\sigma_{+}=0.13$~mag and  $\sigma_{-}=0.18$~mag for $-0.9 < V-K < 0$ (early-type stars, dotted line in the figure). In Fig.~\ref{fig3}c we derive the residual compared to the \citet{2005MNRAS.357..174D} relation (Eq. 2'') which is applicable only in the $-0.1 < V-K < 4$ color domain. We obtain a residual (O-C$_{d}$) which are similar: $\sigma_{+}=0.08$~mag and  $\sigma_{-}=0.07$~mag. This basically means that improving the statistics does not improve the thinnest of the relations. For this purpose, a homogeneous set of V and K photometry is probably required.

We also compare our results with those of \citet{2006A&A...456..789B}, which is, to our knowledge, the only SBC relation, versus V-K, provided for early-type stars in the literature (actually the relation is set from -1.1 to 7, their Table~2), but instead of using Eq.~\ref{equ.1}, they considered another quantity, $\frac{\theta}{9.306.10^{-\frac{V}{5}}}$. We therefore made a conversion to compare with the $S_{\mathrm{V}}$ quantity. The residual  (O-C$_{b}$)  is shown in Fig.~\ref{fig3}d. We find $\sigma_{+}=0.10$~mag and  $\sigma_{-}=0.11$~mag for $0 < V-K <4$ (or late-type stars) and $\sigma_{+}=0.23$~mag and  $\sigma_{-}=-0.23$~mag for $-1 < V-K < 0$ (or early-type stars). These residuals are significantly larger then the ones obtained when using our Eq.~\ref{final} or Eq. 2'' from  \citet{2005MNRAS.357..174D}.

In Fig.~\ref{fig3}a we also have indicated the fast rotating stars and binaries. In Fig.~\ref{fig4} we provide a zoom of the SBC relation over the -1 < V-K < 0.25 color range. In this zoom we have also indicated the uncertainties and the names of the stars in our VEGA sample. We find that the O-C$_{v}$ residual in the (V-K) color range -1~to~0 is $\sigma=0.06$, $\sigma=0.17$, and $\sigma=0.18$, for stars in binary systems (6), for fast rotating stars (8), and for single stars (18). We note the following points:

First, we want to emphasize that a careful selection (by rejected stars with environment and stars with companions in contact), in particular in the range of -1<V-K<0 can significantly improve the precision on the SBC relation. We obtain $\sigma \simeq 0.4$ otherwise.

Second, the dispersion of the O-C residual for stars in binary systems is significantly lower ($0.06$) than to the one obtained with the whole sample (about $0.16$), which indicates that interferometric and photometric measurements are not contaminated by the binarity.

Third, we obtain a large dispersion ($\sigma=0.17$) for fast rotating stars.  Five stars are beyond 1$\sigma$, while three are within, including $\zeta$~Peg in our VEGA sample (see Fig.~\ref{fig3}b). Delta Cygni ($\delta$~Cyg) is, in particular, at 2$\sigma$. In Sect.~\ref{sect3.2}, we estimated the impact of the fast rotation on the angular diameters of $\zeta$~Peg and  $\delta$~Cyg to be $0.039$~mas and $0.047$~mas, respectively. Using Eq.~\ref{equ.1}, it translates into a magnitude effect of $\pm 0.150$~mag and $\pm 0.127$~mag, respectively. As already said, fast rotation modifies several stellar properties such as the shape of the photosphere \citep{1963ApJ...138.1134C,1966ApJ...146..152C} and its brightness distribution \citep{1924MNRAS..84..665V,1924MNRAS..84..684V}, which should be taken into consideration. However, studying these effects requires dedicated modeling, and this will be done in a forthcoming paper. Finally, in our VEGA sample, four stars are beyond 1$\sigma$ from the relation, but when also considering the uncertainty in V-K, they remain consistent with the relation (see Fig.~\ref{fig4}).

Fourth, we calculate the SBC relation for luminosity classes I and II, III, IV, and V (Fig.~\ref{fig33}). We obtain the following results:
\begin{figure}[htbp]
\begin{center}
\includegraphics[width=9cm,height=7cm]{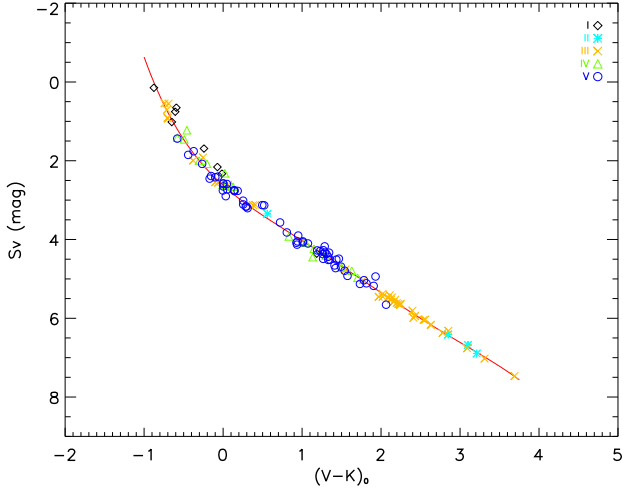}
\end{center}
\caption { The relation between the visual surface brightness $S_{V}$ and the color index $(V-K)_{0}$ for luminosity class I ($\diamond$), luminosity class II ($\ast$),  luminosity class III ($\times$),  luminosity class IV ($\triangle$), and luminosity class V~($\circ$).}
  \label{fig33}
\end{figure}

\begin{eqnarray}
-0.88\leq(V-K)_{0}\leq3.21 \nonumber\\
S_{v}=2.291 + 2.151(V-K)_{0} - 0.461(V-K)^{2}_{0} + 0.073(V-K)^{3}_{0}  \\
\left[\sigma_{S_{v}}=0.08~\mathrm{mag};\sigma_{\theta} \simeq 3.5\%; 12~\mathrm{stars; Class~I+II} \right]\nonumber
\end{eqnarray}

\begin{eqnarray}
-0.74\leq(V-K)_{0}\leq3.69 \nonumber\\
S_{v}=2.497 + 1.916(V-K)_{0} -0.335(V-K)^{2}_{0} + 0.050(V-K)^{3}_{0}\\
\left[\sigma_{S_{v}}=0.07~\mathrm{mag};\sigma_{\theta} \simeq 3.4\%; 41~\mathrm{stars; Class~III} \right]\nonumber
\end{eqnarray}

\begin{eqnarray}
-0.58\leq(V-K)_{0}\leq2.06 \nonumber\\
S_{v}=2.625 + 1.823(V-K)_{0} -0.606(V-K)^{2}_{0} + 0.197(V-K)^{3}_{0}\\
\left[\sigma_{S_{v}}=0.10~\mathrm{mag};\sigma_{\theta} \simeq 4.8\%; 79~\mathrm{stars; Class~IV+V} \right]\cdot\nonumber
\end{eqnarray}

We find a slight difference in the zero-points of these relations. Their dispersion is, however, similar, about 0.09 mag, which is slightly  lower than the global dispersion of 0.16 mag that we obtain when considering the whole sample.

\section{Conclusions}\label{sect6}

Taking advantage of the unique VEGA/CHARA capabilities in terms of spatial resolution, we determined the angular diameters of eight bright early-type stars in the visible with a precision of about 1.5\%.  By combining these data with previous angular diameter determinations, we provide for the very first time a SBC relation for early-type stars with a precision of about 0.16 magnitude, which means that this SBC relation can be used to derive the angular diameter of early-type stars with a precision of 7.3\%. This relation is a powerful tool for the distance scale calibration as it can be used to  derive the individual angular diameters of detached, early-type, and thus bright eclipsing binary systems.  It will be used in the course of the Araucaria Project \citep{2005ApJ...628..695G} to derive the distance of different galaxies in the Local Group, for exemple M33. As the eclipsing binary method is independent of the metallicity of the star, it can be used as a reference to test the impact of the metallicity on several other distance indicators, in particular the Cepheids. In the course of the Araucaria project, we also aim to test the method consistently on galactic early-type eclipsing binaries using photometry, spectroscopy, and interferometry.

\begin{acknowledgements}
This research has made use of the SIMBAD and VIZIER\footnote{Available at http://cdsweb.u- strasbg.fr/} databases at CDS, Strasbourg (France), and of the Jean-Marie Mariotti Center \texttt{Aspro} service\footnote{Available at http://www.jmmc.fr/aspro} and of electronic bibliography maintained by the NASA/ADS system. The research leading to these results has received funding from the European Community's Seventh Framework Programme under Grant Agreement 312430 and financial support from the Ministry of Higher Education and Scientific Research (MHESR) - Tunisia. The CHARA Array is funded by the National Science Foundation through NSF grants AST-0606958 and AST-0908253 and by Georgia State University through the College of Arts and Sciences, as well as the W. M. Keck Foundation. WG gratefully acknowledges financial support for this work from the BASAL Centro de Astrofisica y Tecnologias Afines (CATA) PFB-06/2007, and from the Millenium Institute of Astrophysics (MAS) of the Iniciativa Cientifica Milenio del Ministerio de Economia, Fomento y Turismo de Chile, project IC120009. We acknowledge financial support for this work from ECOS-CONICYT grant C13U01. Support from the Polish National Science Center grant MAESTRO 2012/06/A/ST9/00269 is also acknowledged.
We also wish to thank the referee, Dr Puls, for his numerous and precise suggestions for improving the photometric aspects of the paper. This was an enormous help in refining our results. This research has largely benefited from the support, suggestions, advice of our colleague Olivier Chesneau, who passed away this spring. The whole team wish to pay homage to him.

\end{acknowledgements}

\bibliographystyle{aa}  
\bibliography{challouf} 

\begin{appendix} 

\begin{table*}
\begin{center}
\caption[]{Journal of the observations}
\label{Tabapx1}
\setlength{\doublerulesep}{\arrayrulewidth}
\begin{tabular}{lcccccccc}
\hline \noalign{\smallskip}
\hline\hline\hline\hline
Star  &    Date obs   &   TU  & HA  & MJD    &  $\lambda$ & Base  &  Arg    &  $V^{2}$$\pm_\mathrm{stat}$ $\pm_\mathrm{syst}$ \\
      & [ yyyy-mm-dd ]&  [ h ]&[ h ]&[ days ]&    [ nm ]  & [ m ] & [ deg ] &                                 \\
\hline\hline
$\gamma$ Ori & 2011-10-12   &    $10.206$    &  $-1.707$    &   $55845.5$    &   $710$   &    $65.574$   &   $-114.974$  &    $0.830 \pm 0.059 \pm 0.001 $  \\
 & 2011-10-12   &    $10.209$   &   $-1.704$   &    $55845.5$   &    $731.5$   &    $65.578$   &   $-114.977$   &   $0.817 \pm 0.064 \pm 0.000 $  \\
 & 2011-10-12   &    $10.194$   &   $-1.719$   &    $55845.5$   &    $731.5$   &    $153.578$  &    $-110.390$  &    $0.228 \pm 0.019 \pm 0.001 $  \\
 & 2011-10-13   &    $9.654$    &  $-2.194$    &   $55846.5$    &   $710$   &    $64.439$   &   $-114.627$   &   $0.828 \pm 0.078 \pm 0.000 $  \\
 & 2011-10-13   &    $9.708$    &  $-2.141$    &   $55846.5$    &   $710$   &    $149.824$  &    $-110.214$  &    $0.222 \pm 0.036 \pm 0.001 $  \\
 & 2011-10-13   &    $9.693$    &  $-2.155$    &   $55846.5$    &   $731.5$  &     $149.659$ &     $-110.214$ &     $0.244 \pm 0.024 \pm 0.001 $  \\
 & 2011-11-22   &    $10.851$   &    $1.633$   &    $55886.5$   &    $730$   &    $104.340$   &    $92.583$   &   $0.554 \pm 0.038 \pm 0.001 $  \\
 & 2011-12-10   &    $7.745$    & $-0.2984$    &   $55904.5$    &   $708.5$  &     $154.258$  &    $-112.675$  &    $0.220 \pm 0.012 \pm 0.006 $  \\
\hline
$\gamma$ Lyr & 2011-07-27    &   $5.633$  &   $-0.916$  &    $55768.5$    &   $715$    &   $65.4495$    &  $-116.028$   &   $0.809 \pm 0.060 \pm 0.008 $  \\
  & 2011-07-27   &  $5.633$  &   $-0.916$   &   $55768.5$    &   $715$    &   $154.448$   &   $-109.563$   &   $0.168 \pm 0.020 \pm 0.010 $  \\
  & 2011-07-27   &  $5.633$  &   $-0.916$   &   $55768.5$    &   $735$    &   $65.4495$   &   $-116.028$   &   $0.907 \pm 0.050 \pm 0.009 $  \\
  & 2011-07-27     &  $5.633$    & $-0.916$    &   $55768.5$    &   $735$     &  $154.448$    &  $-109.563$  &    $0.188 \pm 0.014 \pm 0.012 $ \\
  & 2011-07-27     &  $5.633$    & $-0.916$    &   $55768.5$    &   $735$     &  $219.605$    &  $-111.486$  &  $0.002 \pm 0.066 \pm 0.000 $  \\
  & 2011-09-01      & $5.244$     &  $1.060$    &   $55804.5$    &   $715$     &  $63.897$     & $-133.244$   &   $0.699 \pm 0.056 \pm 0.007 $  \\
  & 2011-09-01      & $5.244$     &  $1.060$    &   $55804.5$    &   $715$     &  $152.073$    &  $-126.300$  &    $0.190 \pm 0.026 \pm 0.012 $  \\
  & 2011-09-01      & $5.244$     &  $1.060$    &   $55804.5$    &   $715$     &  $215.640$    &  $-128.353$  &   $0.016 \pm 0.011 \pm 0.002 $  \\
  & 2011-09-01      & $5.244$     &  $1.060$    &   $55804.5$    &   $735$     &  $63.897$     & $-133.244$   &   $0.784 \pm 0.044 \pm 0.007 $  \\
  & 2011-09-01      & $5.244$     &  $1.060$    &   $55804.5$    &   $735$     &  $152.073$    &  $-126.300$  &    $0.222 \pm 0.020 \pm 0.013 $  \\
  & 2011-09-01      & $5.244$     &  $1.060$    &   $55804.5$    &   $735$     &  $215.640$    &  $-128.353$  &  $0.005 \pm 0.012 \pm 0.000 $  \\
  & 2012-06-21     &  $7.755$    &  $-1.103$   &   $56098.5$    &   $707$     &  $65.125$     & $-114.643$   &   $0.659 \pm 0.025 \pm 0.005 $  \\
  & 2012-06-21     &  $7.755$    &  $-1.103$   &    $56098.5$   &    $707$    &   $153.401$   &   $-108.177$  &    $0.239 \pm 0.040 \pm 0.010 $  \\
  & 2012-06-21     &  $7.755$    &  $-1.103$   &    $56098.5$   &    $707$    &  $218.236$    &  $-110.103$  &  $0.026 \pm 0.050 \pm 0.002 $  \\
  & 2012-06-21     &  $7.755$    &  $-1.103$   &    $56098.5$   &    $730.5$   &    $65.125$   &   $-114.643$  &   $0.684 \pm 0.024 \pm 0.004 $  \\
  & 2012-06-21     &  $7.755$    &  $-1.103$   &    $56098.5$   &    $730.5$   &    $153.401$  &    $-108.177$  &    $0.178 \pm 0.017 \pm 0.007 $  \\
  & 2012-06-21     &  $7.755$    &  $-1.103$   &    $56098.5$   &    $730.5$   &    $218.236$  &    $-110.103$  &   $0.046 \pm 0.044 \pm 0.004 $  \\
  & 2012-06-21     &  $9.712$    &  $0.858$    &   $56098.5$    &   $707.5$    &   $64.388$    &  $-131.216$    &  $0.719 \pm 0.044 \pm 0.005 $  \\
  & 2012-06-21     &  $9.712$    &  $0.858$    &   $56098.5$    &   $707.5$    &   $153.274$   &   $-124.351$   &   $0.167 \pm 0.018 \pm 0.007 $  \\
  & 2012-06-21     &  $9.712$    &  $0.858$    &   $56098.5$    &   $707.5$    &   $217.337$   &   $-126.380$  &  $0.003 \pm 0.025 \pm 0.000 $  \\
  & 2012-06-21     &  $9.712$    &  $0.858$    &   $56098.5$    &   $730.5$    &   $64.388$    &  $-131.216$   &  $0.672 \pm 0.046 \pm 0.004 $  \\
  & 2012-06-21     &  $9.712$    &  $0.858$    &   $56098.5$    &   $730.5$    &   $153.274$   &   $-124.351$  &    $0.153 \pm 0.020 \pm 0.006 $  \\
  & 2012-06-21     &  $9.712$    &  $0.858$    &   $56098.5$    &   $730.5$    &   $217.337$   &   $-126.380$  &   $0.043 \pm 0.045 \pm 0.003 $  \\
\hline
$\lambda$ Aql & 2013-07-24    &   $8.216$    &   $1.388$   &   $56496.5$    &   $532.5$    &   $29.264$   &   $-23.993$    &   $1.003\pm0.095\pm0.002$   \\
 & 2013-07-24      & $8.216$     &  $1.388$    &   $56496.5$    &   $532.5$    &   $160.270$   &   $-35.208$   &   $0.205\pm0.020\pm0.023$      \\
 & 2013-07-24      & $8.216$     &  $1.388$    &   $56496.5$    &   $532.5$    &   $189.061$   &   $-33.483$   &   $0.162\pm0.088\pm0.019$  \\
 & 2013-07-24      & $8.216$     &  $1.388$    &   $56496.5$    &   $547.5$    &   $29.264$    &  $-23.993$    &   $1.025\pm0.131\pm0.005$   \\
 & 2013-07-24      & $8.216$     &  $1.388$    &   $56496.5$    &   $547.5$    &   $160.270$   &   $-35.208$   &   $0.201\pm0.015\pm0.029$  \\
 & 2013-07-24      & $8.216$     &  $1.388$    &   $56496.5$    &   $547.5$    &   $189.061$   &   $-33.483$   &   $0.096\pm0.024\pm0.021$  \\
 & 2013-07-24      & $8.925$     &  $2.100$    &   $56496.5$    &   $532.5$    &   $30.595$    &  $-28.275$    &   $0.950\pm0.107\pm0.002$   \\
 & 2013-07-24      & $8.925$     &  $2.100$    &   $56496.5$    &   $532.5$    &   $167.700$   &   $-37.803$   &   $0.142\pm0.010\pm0.012$  \\
 & 2013-07-24      & $8.925$     &  $2.100$    &   $56496.5$    &   $532.5$    &   $197.938$   &   $-36.337$   &   $0.042\pm0.061\pm0.005$  \\
 & 2013-07-24      & $8.925$     &  $2.100$    &   $56496.5$    &   $547.5$    &   $30.595$    &  $-28.275$    &   $0.911\pm0.041\pm0.002$   \\
 & 2013-07-24      & $8.925$     &  $2.100$    &   $56496.5$    &   $547.5$    &   $167.700$   &   $-37.803$   &   $0.169\pm0.012\pm0.014$  \\
 & 2013-07-24      & $8.925$     &  $2.100$    &   $56496.5$    &   $547.5$    &   $197.938$   &   $-36.337$   &   $0.082\pm0.021\pm0.010$  \\
 & 2013-07-24      & $9.319$     &  $2.494$    &   $56496.5$    &   $532.5$    &   $31.311$    &  $-30.138$    &   $0.924\pm0.071\pm0.002$6   \\
 & 2013-07-24      & $9.319$     &  $2.494$    &   $56496.5$    &   $532.5$    &   $171.056$   &   $-38.738$   &   $0.094\pm0.011\pm0.008$  \\
 & 2013-07-24      & $9.319$     &  $2.494$    &   $56496.5$    &   $532.5$    &   $202.070$   &   $-37.410$   &   $0.005\pm0.042\pm0.000$  \\
 & 2013-07-24      & $9.319$     &  $2.494$    &   $56496.5$    &   $547.5$    &   $31.311$    &  $-30.138$    &   $0.928\pm0.050\pm0.002$   \\
 & 2013-07-24      & $9.319$     &  $2.494$    &   $56496.5$    &   $547.5$    &   $171.056$   &   $-38.738$   &   $0.170\pm0.009\pm0.014$  \\
 & 2013-07-24      & $9.319$     &  $2.494$    &   $56496.5$    &   $547.5$    &   $202.070$   &   $-37.410$   &   $0.045\pm0.021\pm0.005$   \\
 & 2013-07-24      & $9.802$     &  $2.979$    &   $56496.5$    &   $532.5$    &   $32.121$    &  $-31.950$    &   $0.928\pm0.105\pm0.002$   \\
 & 2013-07-24      & $9.802$     &  $2.979$    &   $56496.5$    &   $532.5$    &   $174.274$   &   $-39.432$   &   $0.116\pm0.013\pm0.011$  \\
 & 2013-07-24      & $9.802$     &  $2.979$    &   $56496.5$    &   $532.5$    &   $206.164$   &   $-38.270$   &   $0.028\pm0.062\pm0.004$  \\
 & 2013-07-24      & $9.802$     &  $2.979$    &   $56496.5$    &   $547.5$    &   $32.121$    &   $-31.950$   &   $0.998\pm0.090\pm0.002$  \\
 & 2013-07-24      & $9.802$     &  $2.979$    &   $56496.5$    &   $547.5$    &   $174.274$   &   $-39.432$   &   $0.105\pm0.012\pm0.009$  \\
 & 2013-07-24      & $9.802$     &  $2.979$    &   $56496.5$    &   $547.5$    &   $206.164$   &   $-38.270$   &   $0.068\pm0.043\pm0.009$  \\
 & 2013-07-24      & $8.215$     &  $1.388$    &   $56496.5$    &   $703.0$    &   $29.263$    &  $-23.990$    &   $0.972\pm0.235\pm0.001$   \\
 & 2013-07-24      & $8.215$     &  $1.388$    &   $56496.5$    &   $703.0$    &   $160.265$   &   $-35.206$   &   $0.488\pm0.155\pm0.037$  \\
\hline
\end{tabular}
\end{center}
\end{table*}
 \begin{table*}

\begin{center}
\caption{Continued.}
\setlength{\doublerulesep}{\arrayrulewidth}
\label{Tabapx2}
\begin{tabular}{lcccccccc}
\hline\hline
Star  &    Date obs   &   TU  & HA  & MJD    &  $\lambda$ & Base  &  Arg    &  $V^{2}$$\pm_\mathrm{stat}$ $\pm_\mathrm{syst}$ \\
      & [ yyyy-mm-dd ]&  [ h ]&[ h ]&[ days ]&    [ nm ]  & [ m ] & [ deg ] &                                 \\
\hline
 & 2013-07-24      & $8.215$     &  $1.388$    &   $56496.5$    &   $734.0$    &   $29.263$    &  $-23.990$    &   $0.918\pm0.118\pm0.002$   \\
 & 2013-07-24      & $8.215$     &  $1.388$    &   $56496.5$    &   $734.0$    &   $160.265$   &   $-35.206$   &   $0.467\pm0.021\pm0.035$  \\
 & 2013-07-24      & $8.215$     &  $1.388$    &   $56496.5$    &   $734.0$    &   $189.055$   &   $-33.480$   &   $0.236\pm0.068\pm0.026$  \\
 & 2013-07-24      & $8.925$     &  $2.099$    &   $56496.5$    &   $703.0$    &   $30.594$    &  $-28.272$    &   $0.935\pm0.124\pm0.007$   \\
 & 2013-07-24      & $8.925$     &  $2.099$    &   $56496.5$    &   $703.0$    &   $167.695$   &   $-37.802$   &   $0.293\pm0.119\pm0.024$  \\
 & 2013-07-24      & $8.925$     &  $2.099$    &   $56496.5$    &   $734.0$    &   $30.594$    &  $-28.272$    &   $0.895\pm0.075\pm0.001$   \\
 & 2013-07-24      & $8.925$     &  $2.099$    &   $56496.5$    &   $734.0$    &   $167.695$   &   $-37.802$   &   $0.399\pm0.015\pm0.017$  \\
 & 2013-07-24      & $8.925$     &  $2.099$    &   $56496.5$    &   $734.0$    &   $197.932$   &   $-36.335$   &   $0.434\pm0.146\pm0.025$  \\
 & 2013-07-24      & $9.318$     &  $2.494$    &   $56496.5$    &   $703.0$    &   $31.309$    &  $-30.134$    &   $0.910\pm0.269\pm0.002$   \\
 & 2013-07-24      & $9.318$     &  $2.494$    &   $56496.5$    &   $703.0$    &   $171.050$   &   $-38.736$   &   $0.431\pm0.115\pm0.020$  \\
 & 2013-07-24      & $9.318$     &  $2.494$    &   $56496.5$    &   $734.0$    &   $31.309$    &  $-30.134$    &   $0.895\pm0.075\pm0.001$   \\
 & 2013-07-24      & $9.318$     &  $2.494$    &   $56496.5$    &   $734.0$    &   $171.050$   &   $-38.736$   &   $0.358\pm0.015\pm0.016$  \\
 & 2013-07-24      & $9.318$     &  $2.494$    &   $56496.5$    &   $734.0$    &   $202.062$   &   $-37.408$   &   $0.550\pm0.279\pm0.062$  \\
 & 2013-07-24      & $9.801$     &  $2.979$    &   $56496.5$    &   $703.0$    &   $32.120$    &  $-31.948$    &   $0.936\pm0.090\pm0.002$   \\
 & 2013-07-24      & $9.801$     &  $2.979$    &   $56496.5$    &   $703.0$    &   $174.269$   &   $-39.431$   &   $0.369\pm0.189\pm0.018$  \\
 & 2013-07-24      & $9.801$     &  $2.979$    &   $56496.5$    &   $703.0$    &   $206.158$   &   $-38.269$   &   $0.885\pm0.394\pm0.054$  \\
 & 2013-07-24      & $9.801$     &  $2.979$    &   $56496.5$    &   $734.0$    &   $32.120$    &  $-31.948$    &   $1.122\pm0.100\pm0.002$   \\
 & 2013-07-24      & $9.801$     &  $2.979$    &   $56496.5$    &   $734.0$    &   $174.269$   &   $-39.431$   &   $0.367\pm0.015\pm0.017$  \\
 & 2013-07-24      & $9.801$     &  $2.979$    &   $56496.5$    &   $734.0$    &   $206.158$   &   $-38.269$   &   $0.318\pm0.089\pm0.036$  \\
\hline
$\iota$ Her & 2013-08-29    &   $4.351$    &   $1.325$   &   $56532.5$    &   $538.5$    &   $106.359$   &   $-94.209$    &   $0.875\pm0.048\pm0.006$   \\
 & 2013-08-29      & $4.351$     &  $1.325$    &   $56532.5$    &   $538.5$    &   $310.123$   &   $-120.677$  &   $ 0.081\pm0.056\pm0.004$  \\
 & 2013-08-29      & $4.351$     &  $1.325$    &   $56532.5$    &   $553.5$    &   $106.359$   &   $-94.209$   &   $ 0.930\pm0.052\pm0.006$  \\
 & 2013-08-29      & $4.351$     &  $1.325$    &   $56532.5$    &   $553.5$    &   $310.123$   &   $-120.677$  &   $ 0.118\pm0.024\pm0.007$  \\
 & 2013-08-29      & $4.350$     &  $1.324$    &   $56532.5$    &   $707.5$    &   $106.361$   &   $-94.203$   &   $ 0.933\pm0.122\pm0.004$  \\
 & 2013-08-29      & $4.350$     &  $1.324$    &   $56532.5$    &   $707.5$    &   $310.127$   &   $-120.670$  &   $ 0.430\pm0.164\pm0.015$  \\
 & 2013-08-29      & $4.350$     &  $1.324$    &   $56532.5$    &   $738.5$    &   $106.361$   &   $-94.203$   &   $ 0.904\pm0.035\pm0.004$  \\
 & 2013-08-29      & $4.350$     &  $1.324$    &   $56532.5$    &   $738.5$    &   $310.127$   &   $-120.670$  &   $ 0.315\pm0.054\pm0.010$  \\
\hline
$8$ Cyg & 2013-08-28    &   $6.088$    &   $1.130$   &   $56531.5$    &   $538.5$    &   $216.020$   &   $-129.391$    &   $0.630\pm0.053\pm0.035$   \\
 & 2013-08-28      & $6.088$     &  $1.130$    &   $56531.5$    &   $538.5$    &   $306.627$   &   $-116.749$  &   $0.407\pm0.108\pm0.049$   \\
 & 2013-08-28      & $6.088$     &  $1.130$    &   $56531.5$    &   $553.5$    &   $216.020$   &   $-129.391$  &   $0.617\pm0.046\pm0.033$   \\
 & 2013-08-28      & $6.088$     &  $1.130$    &   $56531.5$    &   $553.5$    &   $306.627$   &   $-116.749$  &   $0.412\pm0.079\pm0.047$   \\
 & 2013-08-28      & $6.088$     &  $1.129$    &   $56531.5$    &   $707.5$    &   $216.025$   &   $-129.386$  &   $0.776\pm0.053\pm0.024$   \\
 & 2013-08-28      & $6.088$     &  $1.129$    &   $56531.5$    &   $707.5$    &   $306.634$   &   $-116.743$  &   $0.442\pm0.080\pm0.029$   \\
 & 2013-08-28      & $6.088$     &  $1.129$    &   $56531.5$    &   $740.0$    &   $216.025$   &   $-129.386$  &   $0.973\pm0.151\pm0.028$   \\
 & 2013-08-28      & $6.088$     &  $1.129$    &   $56531.5$    &   $740.0$    &   $306.634$   &   $-116.743$  &   $0.431\pm0.074\pm0.025$   \\
\hline
$\zeta$ Per & 2011-10-13  &  $8.507$ &  $-1.828$  &  $55846.5$  &  $715.0$    &   $63.035$    &  $-109.781$   &   $0.965\pm0.051\pm0.001$   \\
 & 2011-10-13     & $8.507$     &  $-1.828$   &   $55846.5$    &   $715.0$    &   $147.169$   &   $-103.253$  &   $0.471\pm0.028\pm0.004$    \\
 & 2011-10-13     & $8.484$     &  $-1.851$   &   $55846.5$    &   $715.0$    &   $209.574$   &   $-105.057$  &   $0.181\pm0.014\pm0.003$    \\
 & 2011-10-13     & $8.507$     &  $-1.828$   &   $55846.5$    &   $734.5$    &   $63.035$    &   $-109.781$  &   $0.904\pm0.032\pm0.001$     \\
 & 2011-10-13     & $8.507$     &  $-1.828$   &   $55846.5$    &   $734.5$    &   $147.169$   &   $-103.253$  &   $0.494\pm0.016\pm0.004$     \\
 & 2011-10-13     & $8.500$     &  $-1.835$   &   $55846.5$    &   $734.5$    &   $209.811$   &   $-105.163$  &   $0.228\pm0.012\pm0.004$     \\
\hline\hline
\end{tabular}
\end{center}
\end{table*}

\begin{table*}
\begin{center}
\caption[]{Continued}
\label{Tabapx3}
\setlength{\doublerulesep}{\arrayrulewidth}
\begin{tabular}{lcccccccc}
\hline \noalign{\smallskip}
\hline\hline\hline\hline
Star  &    Date obs   &   TU  & HA  & MJD    &  $\lambda$ & Base  &  Arg    &  $V^{2}$$\pm_\mathrm{stat}$ $\pm_\mathrm{syst}$ \\
      & [ yyyy-mm-dd ]&  [ h ]&[ h ]&[ days ]&    [ nm ]  & [ m ] & [ deg ] &                                 \\
\hline\hline
$\delta$ Cyg & 2011-07-23    &   $7.421$  &   $-0.153$    &   $55764.5$   &    $715$   &    $65.642$    &  $-121.782$   &   $0.772 \pm 0.041 \pm 0.002$     \\
  & 2011-07-23    &   $7.425$  &   $-0.149$    &   $55764.5$   &    $715$   &    $155.612$    &  $-115.029$    &  $0.166 \pm 0.011 \pm 0.002 $  \\
  & 2011-07-23    &   $7.421$  &   $-0.153$    &   $55764.5$   &    $715$   &    $65.642$     &  $-121.782$    &  $0.772 \pm 0.040 \pm 0.002 $  \\
  & 2011-07-23    &   $7.425$  &   $-0.149$    &   $55764.5$   &    $715$   &    $155.612$    &  $-115.029$    &  $0.166 \pm 0.010 \pm 0.002 $  \\
  & 2011-07-23    &   $7.421$  &   $-0.153$    &   $55764.5$   &    $735$   &    $65.642$    &  $-121.782$    &  $0.698 \pm 0.035 \pm 0.001 $  \\
  & 2011-07-23    &   $7.419$  &   $-0.155$    &   $55764.5$   &    $735$   &    $155.596$    &  $-114.963$    &  $0.147 \pm 0.008 \pm 0.002 $  \\
  & 2011-07-23    &   $9.054$  &   $1.484$     &   $55764.5$   &    $715$   &    $65.146$    &  $-139.701$    &  $0.708 \pm 0.059 \pm 0.001 $  \\
  & 2011-07-23    &   $9.050$  &   $1.480$     &   $55764.5$   &    $715$   &    $154.249$    &  $-132.820$    &  $0.127 \pm 0.012 \pm 0.001 $  \\
  & 2011-07-23    &   $9.054$  &   $1.484$     &   $55764.5$   &    $715$   &    $219.055$    &  $-134.898$    &  $0.011 \pm 0.019 \pm 0.003 $  \\
  & 2011-07-23    &   $9.054$  &   $1.484$     &   $55764.5$   &    $735$   &    $65.146$    &  $-139.701$    &  $0.684 \pm 0.040 \pm 0.001 $  \\
  & 2011-07-23    &   $9.048$  &   $1.478$     &   $55764.5$   &    $735$   &    $154.256$    &  $-132.797$    &  $0.160 \pm 0.009 \pm 0.002 $  \\
  & 2011-07-23    &   $9.054$  &   $1.484$     &   $55764.5$   &    $735$   &    $219.055$    &  $-134.898$    &  $0.016 \pm 0.010 \pm 0.000 $  \\
  & 2011-07-27    &   $8.124$  &   $0.815$     &   $55768.5$   &    $715$   &    $65.744$    &  $-132.043$    &  $0.726 \pm 0.040 \pm 0.001 $  \\
  & 2011-07-27    &   $8.124$  &   $0.815$     &   $55768.5$   &    $715$   &    $155.931$    &  $-125.241$    &  $0.128 \pm 0.013 \pm 0.002 $  \\
  & 2011-07-27    &   $8.124$  &   $0.815$     &   $55768.5$   &    $735$   &    $65.7443$    &  $-132.043$    &  $0.788 \pm 0.049 \pm 0.002 $  \\
  & 2011-07-27    &   $8.124$  &   $0.815$     &   $55768.5$   &    $735$   &    $155.931$    &  $-125.241$    &  $0.157 \pm 0.014 \pm 0.002 $  \\
  & 2011-07-27    &   $8.124$  &   $0.815$     &   $55768.5$   &    $735$   &    $221.349$    &  $-127.257$    &  $0.028 \pm 0.019 \pm 0.000 $  \\
  & 2011-07-27    &   $8.917$  &   $1.610$     &   $55768.5$   &    $715$   &    $65.007$    &  $-141.195$    &  $0.770 \pm 0.073 \pm 0.001 $  \\
  & 2011-07-27    &   $8.917$  &   $1.610$     &   $55768.5$   &    $715$   &    $153.819$    &  $-134.361$    &  $0.163 \pm 0.015 \pm 0.002 $  \\
  & 2011-07-27    &   $8.917$  &   $1.610$     &   $55768.5$   &    $715$   &    $218.501$    &  $-136.390$    &  $0.018 \pm 0.020 \pm 0.000 $  \\
  & 2011-07-27     &  $8.917$   &    $1.610$    &   $55768.5$   &    $735$   &    $65.007$     & $-141.195$     & $0.797 \pm 0.053 \pm 0.001 $  \\
  & 2011-07-27     &  $8.917$   &    $1.610$    &   $55768.5$   &    $735$   &    $153.819$    &  $-134.361$    &  $0.160 \pm 0.012 \pm 0.002 $  \\
\hline
\hline
$\zeta$ Peg & 2011-07-24   &    $8.586$   &   $-1.860$    &   $55765.5$    &   $715$   &   $65.002$   &   $-114.225$   &   $0.867 \pm 0.039 \pm 0.005 $  \\
 & 2011-07-24   &    $8.583$   &   $-1.863$   &    $55765.5$   &    $715$   &    $151.533$  &    $-109.301$   &   $0.428 \pm 0.021 \pm 0.015 $  \\
 & 2011-07-24   &    $8.576$   &   $-1.870$   &    $55765.5$   &    $715$   &    $216.288$  &    $-110.765$   &   $0.263 \pm 0.028 \pm 0.022 $  \\
 & 2011-07-24   &    $8.586$   &   $-1.860$   &    $55765.5$   &    $735$   &    $65.002$   &   $-114.225$    &  $0.839 \pm 0.034 \pm 0.005 $  \\
 & 2011-07-24   &    $8.583$   &   $-1.863$   &    $55765.5$   &    $735$   &    $151.533$  &    $-109.301$   &   $0.414 \pm 0.015 \pm 0.014 $  \\
 & 2011-07-24   &    $8.589$   &   $-1.857$   &    $55765.5$   &    $735$   &    $216.431$  &    $-110.787$   &   $0.222 \pm 0.017 \pm 0.015 $  \\
 & 2011-07-28   &    $8.236$   &   $-1.948$   &    $55769.5$   &    $715$   &    $64.774$   &   $-114.068$    &  $0.891 \pm 0.059 \pm 0.054 $  \\
 & 2011-07-28   &    $8.236$   &   $-1.948$   &    $55769.5$   &    $715$   &    $150.724$  &    $-109.167$   &   $0.442 \pm 0.027 \pm 0.014 $  \\
 & 2011-07-28   &    $8.213$   &   $-1.971$   &    $55769.5$   &    $715$   &    $215.043$  &    $-110.606$   &   $0.187 \pm 0.028 \pm 0.013 $  \\
 & 2011-07-28   &    $8.240$   &   $-1.944$   &    $55769.5$   &    $735$   &    $64.784$   &   $-114.074$    &  $0.882 \pm 0.055 \pm 0.005 $  \\
 & 2011-07-28   &    $8.240$   &   $-1.944$   &    $55769.5$   &    $735$   &    $150.760$  &    $-109.173$   &   $0.440 \pm 0.030 \pm 0.013 $  \\
 & 2011-07-28   &    $8.243$   &   $-1.941$   &    $55769.5$   &    $735$    &   $215.419$   &   $-110.650$    &  $0.188 \pm 0.045 \pm 0.012 $  \\
\hline\hline
\end{tabular}
\end{center}
\end{table*}

\end{appendix}

\end{document}